\documentclass[12pt,letter,doublespace]{article}
\usepackage{amssymb}
\usepackage{natbib,graphicx,setspace,lscape,longtable}
\usepackage{natbib,epsfig,graphicx}
\usepackage{amsmath,amsthm,amssymb}
\usepackage{algorithmic}
\usepackage{color,xcolor}
\usepackage{threeparttable}
\usepackage{changes}

\usepackage{booktabs}

\usepackage{hyperref}

\usepackage[linesnumbered,boxed,ruled,commentsnumbered]{algorithm2e}

\RequirePackage[mathlines, displaymath]{lineno}
\usepackage{multirow}
\usepackage{graphicx}
\usepackage{subfig}
\usepackage{threeparttable}
\DeclareGraphicsExtensions{.eps, .jpg, .pdf}
\bibpunct{(}{)}{;}{a}{,}{,}

\newcommand{\csection}[1]
    {\begin{center}
        \stepcounter{section}
        {\bf\large\arabic{section}. #1}
    \end{center}
    \vspace{-0.15 cm}
}

\newcommand{\scsection}[1]
    {\begin{center}
        {\bf\large #1}
    \end{center}
    \vspace{-0.15 cm}
}

\def\beqr{\begin{eqnarray}}
\def\eeqr{\end{eqnarray}}
\def\beqrs{\begin{eqnarray*}}
\def\eeqrs{\end{eqnarray*}}
\def\bep{\begin{prop}}
\def\eep{\end{prop}}
\def\bc{\begin{center}}
\def\ec{\end{center}}

\newcommand{\bs}{\mathbf{s}}

\newcommand{\Z}{\mathbf{Z}}
\newcommand{\U}{\mathbf{U}}
\newcommand{\C}{\mathbf{C}}
\newcommand{\bH}{\mathbf{H}}
\newcommand{\D}{\mathbf{D}}
\newcommand{\V}{\mathbf{V}}
\newcommand{\I}{\mathbf{I}}
\newcommand{\M}{\mathbf{M}}
\newcommand{\bP}{\mathbf{P}}

\newcommand{\trans}{^{\mbox{\tiny{T}}}}

\numberwithin{equation}{section}

\def \bec{\begin{center}}
\def \enc {\end{center}}
\def \bee {\begin{eqnarray*}}
\def \ene {\end{eqnarray*}}

\def \bear{\begin{array}}
\def \enar{\end{array}}

\def \bs{\begin{slide}}
\def \es{\end{slide}}

\newcommand{\bbeta}{\boldsymbol{\beta}}
\newcommand{\aalpha}{\boldsymbol{\alpha}}
\newcommand{\ggamma}{\boldsymbol{\gamma}}
\newcommand{\ttheta}{\boldsymbol{\theta}}

\newcommand{\SSigma}{\boldsymbol{\Sigma}}
\newcommand{\ddelta}{\boldsymbol{\delta}}
\newcommand{\pphi}{\boldsymbol{\phi}}
\newcommand{\xxi}{\boldsymbol{\xi}}

\newcommand{\PPsi}{\boldsymbol{\Psi}}
\newcommand{\eeta}{\boldsymbol{\eta}}

\newcommand{\LLambda}{\boldsymbol{\Lambda}}

\newcommand{\bZ}{{\bf Z}}
\newcommand{\bW}{{\bf W}}
\newcommand{\bS}{{\bf S}}

\numberwithin{equation}{section}  

\newtheorem{thm}{Theorem}[section]
\newtheorem{lem}{Lemma}[section]

\newtheorem{prop}{Proposition}[section]

\renewcommand{\baselinestretch}{1.2}
\begin{document}

\title{ Estimation and Inference for Multi-Kink Quantile Regression}
\author{
Xiamen University, China}

\author{{Wei Zhong$^1$, Chuang Wan$^{1}$ and Wenyang Zhang$^{2}$}\\
Xiamen University$^1$ and The University of York$^2$}

\maketitle{}

\begin{abstract}
The Multi-Kink Quantile Regression (MKQR) model is an important tool for
analyzing data with heterogeneous conditional distributions, especially when
quantiles of response variable are of interest, due to its robustness to outliers and heavy-tailed errors in the response.
It assumes different linear quantile regression forms in different regions
of the domain of the threshold covariate but are still continuous at kink points.
In this paper, we investigate parameter estimation, kink point detection and statistical
inference in MKQR models.  We propose an iterative segmented quantile
regression algorithm for estimating both the regression coefficients and the
locations of kink points.  The proposed algorithm is much more computationally
efficient than the grid search algorithm and not sensitive to the selection of
initial values.  Asymptotic properties, such as selection consistency of
the number of kink points, asymptotic normality of the estimators of both
regression coefficients and kink effects,
are established to justify the proposed method theoretically.  A score test,
based on partial subgradients, is developed to verify whether the kink effects
exist or not. Test-inversion confidence intervals for kink location parameters are
also constructed.  Intensive simulation studies conducted show the proposed methods
work very well when sample size is finite.  Finally, we apply the MKQR models
together with the proposed methods to the dataset about secondary industrial
structure of China and the dataset about triceps skinfold thickness of Gambian
females, which leads to some very interesting findings.  A new R package
\emph{MultiKink} is developed to implement the proposed methods.
\end{abstract}

\noindent{\bf Keywords}:  Change point detection, hypothesis testing, kink regression, model selection, quantile regression.


\newpage
\pagestyle{plain}
\setcounter{page}{1}
\csection{INTRODUCTION}
The linear models are the most commonly used models in data analysis, however,
the assumption of linearity on the relationship between response variable and
covariates often does not hold in reality.  Fully nonparametric modelling may
suffer from ``curse of dimensionality" and lack of interpretability,
and some important information may not be used even if it is available.
In data analysis, sometimes, some information
about the shape of the underlying model is available, for example,
\cite{li2011bent} proposed a bent line regression with one threshold point
and showed that the logarithm of maximal running speed of land
mammals linearly increases with the logarithm of mass up to a certain point and
then decreases as the mass rises.  Also, in our empirical study on secondary
industrial structure of China, we find, see Figure \ref{fig:3}, that the
city-level proportion of the secondary industry increases quickly with the GDP
per capita up to a certain threshold around 5000-6500 US dollars and then
stabilizes with a slow increasing rate.  Ignoring such information would
eventually pay a price on variance side of the final estimators.  Appealing
more flexible parametric models by making use of the information about the
shape of the underlying model would be a useful approach in data analysis.

The kink regression models \citep{hansen2017regression},
also referred to as bent line regression \citep{li2011bent} or continuous threshold regression,
assume linear regression forms are separately modeled on two sides of an unknown threshold but still
continuous at the threshold.  It is a very useful tool to deal with nonlinearity in data analysis.  Let $Y_t$ be a response variable of interest, $X_t$ a univariate threshold variable and
$\bZ_t$ a $p$ dimensional random vector of additional covariates, $t=1, \ 2, \
\cdots, \ n$.  \cite{hansen2017regression} considered the following kink
regression model with an unknown threshold,
\begin{eqnarray}\label{eq:basic}
Y_t=\alpha_0+\alpha_1(X_t-\delta)I(X_t\leq\delta)+\alpha_2(X_t-\delta)I(X_t>\delta)
+\ggamma\trans\bZ_t+e_t,
\end{eqnarray}
where $e_t$ is the random error with
$E(e_t | X_t, \ \bZ_t) = 0$.
In Model (\ref{eq:basic}), the threshold variable $X_t$ has different slopes on
different segments formed by $\delta$, and the regression function is
continuous with respect to $X_t$.  This kind of non-linear pattern is commonly
referred to as kink effect \citep{hansen2017regression} or bent line effect
\citep{li2011bent}.  The parameter $\delta$ is therefore called
``change point'', ``kink point'' or ``threshold'' exchangeably to represent the
point where the regression function form changes.
Compared with the linear regression models, kink regression models relax the
linearity assumption, therefore, are able to capture the necessary nonlinearity
and make the models more flexible in applications.  Compared with the
nonparametric modelling, kink regression models have the better
interpretability by maintaining linear regression models in different regions
of the domain of $X_t$.  Thus, kink regression models enjoy the
interpretability of linear models as well as the flexibility of nonparametric
regression models.

In the literature, the kink regression models with a single unknown threshold
point have been intensively studied.  For example, \cite{li2011bent} proposed
a continuous bent line quantile regression model and discussed three methods
for testing the existence of a change point.  \cite{hansen2017regression}
combined the least squares estimation and a grid search algorithm to estimate the
regression coefficients and the kink point.  The F-type statistic test was also
proposed for testing $H_0: \alpha_1=\alpha_2$ in Model (\ref{eq:basic}).  In the application,
\cite{hansen2017regression} demonstrated the famous inverted U-shaped
relationship between the GDP growth rate and the ratio of debt to GDP
\citep{reinhart2010growth}.
\cite{zhang2017continuous} studied estimation and hypothesis testing for a
continuous threshold expectile model.  \cite{fong2019fast} developed fast
bootstrap confidence intervals for continuous threshold linear regression.
\cite{hidalgo2019robust} discussed whether there experiences  a discontinuous
jump or a continuous kink at the threshold point by using the quasi-likelihood-ratio
test and  constructed a robust confidence interval for the threshold.

Apparently the kink regression models with one threshold point are not
sufficient in some applications.  In our second real data example, the logarithm of
triceps skinfold thickness (TSF) as an important measure of body density
decreases with the age in the childhood up to about 10 years old, then
experiences a growth spurt at adolescence up to about 18-20 years old and
finally stays almost stable for adults  (see Figure \ref{fig:4}).  The kink
regression models with one threshold are clearly not appropriate for this
case.  It would be more sensible to consider the kink regression models with
multiple threshold points. Moreover,
To achieve the robustness to outliers and heavy-tailed errors which are often present in the data, we consider quantile regression to analyze data with heterogeneous conditional distributions. Quantile regression is able to provide a comprehensive picture of the conditional distribution of the response $Y_t$ given the covariates, especially when its upper or lower quantiles are particularly of interest. Thus, we are motivated to consider a new Multi-Kink Quantile Regression (MKQR) model with $K$ unknown kink points, for a given quantile level $\tau\in (0,1)$,
\begin{eqnarray}\label{eq:mk}
Q_Y(\tau|X_t,\bZ_t)
=\left\{
\begin{array}{ll}
\alpha_0^{(1)}+\alpha_1^{(1)}X_t+\ggamma\trans\bZ_t, & \text{if $X_t\leq\delta_1$}\\
\alpha_0^{(2)}+\alpha_1^{(2)}X_t+\ggamma\trans\bZ_t, & \text{if $\delta_1<X_t\leq\delta_2$}\\
\vdots & \vdots\\
\alpha_0^{(K)}+\alpha_1^{(K)}X_t+\ggamma\trans\bZ_t, &\text{if $\delta_{K-1}<X_t\leq\delta_K$}\\
\alpha_{0}^{(K+1)}+\alpha_{1}^{(K+1)}X_t+\ggamma\trans\bZ_t, & \text{if $X_t>\delta_K$}
\end{array}\right.\quad t=1,\cdots,n,
\\[3mm]
\text{subject to} \quad \alpha_0^{(k)}+\alpha_1^{(k)}\delta_{k}=\alpha_0^{(k+1)}+\alpha_1^{(k+1)}\delta_k,\quad\text{for}\quad k=1,\cdots,K,\nonumber
\end{eqnarray}
where $Q_Y(\tau|X_t,\bZ_t)$ is the $\tau$th conditional quantile of $Y_t$ given $X_t$ and $\bZ_t$, $\alpha_0^{(k)}$ and $\alpha_1^{(k)}$ are the intercept and the slope of $X_t$ in the $k$th segment for $k=1, \cdots,K+1$, respectively, and $\ggamma$ is the coefficient vector of covariates $\bZ_t$, which stay constant on the whole domain of $\bZ_t$,
$\{\delta_k,k=1,\cdots,K\}$ represent the kink points or the locations where kink effects happen satisfying $\delta_1<\cdots<\delta_{K}$,
and the constraints are imposed to ensure the continuity of the regression curve at kink points. 
Thus, there are $K+1$ regimes in total divided by $K$ kink points.  The number of kink points $K$ and their locations are both unknown. Note that all the unknown parameters depend on the quantile index $\tau$, but we omit the subscript $\tau$ for ease of notations throughout the paper.

The MKQR model (\ref{eq:mk}) brings some technical challenges to both
parameter estimation and statistical inference.
From a computational perspective, the grid search
approach commonly used in single kink regression models is no longer practicable especially when $K$ is large, because its computational cost grows at an exponential rate of $K$. We remark that
the MKQR model (\ref{eq:mk}) is different from traditional threshold regression models  that specify
different regression functions in subsamples segmented by a continuous threshold variable in the literature \citep{hansen2000threshold,caner2002threshold,li2011threshold,zhang2014testing}.
The threshold regression models can be viewed as special cases of
varying coefficient models, where the threshold variable is generally not an predictor in the regression.
However, the threshold variable $X_t$ in the MKQR model is also the predictor of interest in the regression. Meanwhile, the MKQR model requires that the regression curves are everywhere continuous on the whole domain of the threshold variable $X_t$, while jumps at the threshold points are allowed in the threshold regression models.

In this paper, we focus on parameter estimation, kink points detection and statistical inference for the MKQR model (\ref{eq:mk}) where the number of kink points and their locations are both unknown using  quantile regression. We contribute the literature in the following several aspects. First, the MKQR model (\ref{eq:mk}) extends the existing kink regression with an unknown threshold to 
wider applications with unknown multiple kink points. We propose a Bootstrap Restarting Iterative Segmented Quantile (BRISQ) regression algorithm for estimating both the regression coefficients and kink effects. This algorithm is much more computationally efficient than the grid search algorithm and not sensitive to the initial values due to the bootstrap restarting idea of \cite{wood2001minimizing}. Furthermore, we suggest a backward elimination algorithm to identify the number of kinks by transforming change points  detection into model selection problem based on quantile BIC criteria.
Second, we theoretically demonstrate that the selection consistency of the number of kink points and the
asymptotical normality of the estimators for both regression coefficients and kink effects.
Third, the MKQR model is robust to outliers in the response and heavy-tailed errors and more flexible for modelling data with heterogeneous conditional distributions especially when upper or lower quantiles of the response are particularly of interest.
Forth, from the statistical inference perspective, we develop a score test based on partial subgradient of quantile objective function under the null hypothesis to verify whether the kink effects exist or not. A test-inversion confidence interval based on a smoothed rank score test for a kink location parameter is also proposed and can be extended to multiple kink parameters by sample splitting.
Fifth, two real data on secondary industrial structure of China and triceps skinfold thickness
 for Gambian females are studied to identify the kink points which would be of interest for economists and biologists, respectively. Last, a new R package \emph{MultiKink}  is developed to implement all the estimation and inference procedures, and is free to use.

The rest of the paper is structured as follows. In Section 2, we describe the estimation procedures
for the MKQR model and investigate the asymptotic properties. Section 3 presents a testing procedure for the existence of kink effects and construct the test-inversion confidence intervals for the kink location parameters. The finite sample performances of the proposed methods are evaluated via simulation experiments in Section 4. In Section 5, two real data applications are studied to illustrate the proposed methodologies. Section 6  concludes the remarks. The technique proofs are presented in the Appendix.

\csection{ESTIMATION AND ALGORITHM}
\vspace{-.5cm}
\subsection{Parameter Estimation}
 Since the original multiple kink model (\ref{eq:mk}) is too complex with many redundant parameters and the continuity constraints, we reparameterize it as
 \begin{equation}\label{eq:mkqr}
Q_Y(\tau|X_t,\bZ_t)=\alpha_0+\alpha_1X_t+\sum_{k=1}^K\beta_k(X_t-\delta_k)I(X_t>\delta_k)+\ggamma\trans\bZ_t,
\end{equation}
where $\alpha_0=\alpha_0^{(1)}$, $\alpha_1=\alpha_1^{(1)}$ and $\beta_k=\alpha_1^{(k+1)}-\alpha_1^{(k)}$, which represents the  difference in slopes for $X_t$ between two adjacent segments.
Thus, $\beta_k\neq0$ implies the existence of a kink effect at $X_t=\delta_k$.
For notation convenience, we let $\bbeta=(\beta_{1},\cdots,\beta_{K})\trans$, $\eeta=(\alpha_{0},\alpha_{1},\bbeta\trans,\ggamma\trans)\trans$ and $\ddelta=(\delta_{1},\cdots,\delta_{K})\trans$. Notice that both the dimensions of $\eeta$ and $\ddelta$ hinge on $K$.

Let $\bW_t=(X_t,\bZ_t\trans)\trans$ and $\ttheta=(\eeta\trans,\ddelta\trans)\trans$.
We rewrite the $\tau$th conditional quantile of $Y_t$ given $\bW_t$ as $Q_Y(\tau;\ttheta|\bW_t)$ to emphasize the dependence on the parameter $\ttheta$.
To estimate $\ttheta=(\eeta\trans,\ddelta\trans)\trans$, we define the following objective function,
\begin{equation}\label{eq:obj}
S_n(\ttheta)=n^{-1}\sum_{t=1}^n\rho_\tau\{Y_t-Q_{Y}(\tau;\ttheta|\bW_t)\},
\end{equation}
where $\rho_\tau(u)=u\{\tau-I(u<0)\}$. A common estimator for $\ttheta$ is thereby
\begin{equation}\label{eq:est}
\widehat{\ttheta}=(\widehat{\eeta}\trans,\widehat{\ddelta}\trans)\trans=\underset{\eeta\in\mathcal{B},\ddelta\in\Gamma}{\arg\min}
S_n(\ttheta),
\end{equation}
where $\mathcal{B}\subset\mathbb{R}^{2+K+p}$ and  $\Gamma\subset\Omega^K$ are compact sets, in which  $\Omega$ denotes the support of threshold variable $X_t$.
If all kink locations parameters $\ddelta$ are known, a simple quantile regression can be directly used for (\ref{eq:mkqr}).
For the single kink regression with an unknown threshold, the greedy grid search algorithm \citep{li2011bent, hansen2017regression} can be used to exhaustively seek the kink point. However, without any prior information, we have to assume that both the number of kink points $K$ and the kink locations vector $\ddelta$ are unknown in Model (\ref{eq:mkqr}). It makes the grid search approach inappropriate especially when $K$ is large, because its computational cost grows at an exponential rate of $K$.
Since the minimization problem in (\ref{eq:obj}) is non-convex with respect to $\ddelta$, the convex optimization algorithms can not directly applied. To this end, we develop a new iterative segmented quantile regression algorithm to detect the number of kink points and estimate regression coefficients simultaneously.

\subsubsection{Bootstrap Restarting Iterative Segmented Quantile Algorithm} 
We  begin our algorithm with a fixed $K$. Although $K$ is given, $(X-\delta_k)I(X>\delta_k)$ for $k=1,\cdots,K$ are not observable and still non-differentiable at $\delta_k$'s. Given an initial location vector $\ddelta^{(0)}=(\delta_1^{(0)},\cdots,\delta_K^{(0)})\trans$, we employ the first-order Taylor expansion to approximate the nonlinear term $(X_t-\delta_k)I(X_t>\delta_k)$ around $\delta^{(0)}_k$,
$$
(X_t-\delta_k)I(X>\delta_k)\approx(X_t-\delta_k^{(0)})I(X_t>\delta_k^{(0)})-(\delta_k-\delta_k^{(0)})I(X_t>\delta_k^{(0)}).
$$
 Then, Model (\ref{eq:mkqr}) can be approximated by
\begin{equation}\label{eq:6}
Q_Y(\tau;\ttheta|\bW_t)\approx\alpha_0+\alpha_1X_t+\sum_{k=1}^K\beta_k\widetilde{U}_{kt}+
\sum_{k=1}^K\phi_k\widetilde{V}_{kt}+\ggamma\trans\bZ_t,
\end{equation}
where $\phi_k=\beta_k(\delta_k-\delta_k^{(0)})$, $\widetilde{U}_{kt}=(X_t-\delta_k^{(0)})I(X_t>\delta_k^{(0)})$ and $\widetilde{V}_{kt}=-I(X_t>\delta_k^{(0)})$ are two new covariates with coefficients $\beta_k$ and $\phi_k$, respectively. Denote $\bbeta=(\beta_1,\cdots,\beta_K)\trans$ and $\pphi=(\phi_1,\cdots,\phi_K)\trans$.
This local linear approximation technique has been also used in the change point detection, such as \cite{muggeo2010efficient} for the piecewise constant model.
By fitting the standard  linear quantile model (\ref{eq:6}), a new estimator for ${\delta}_k$ can be updated by $\widehat{\delta}_k^{(1)}={\delta}_k^{(0)}+\widehat{\phi}_k/\widehat{\beta}_k$, for $k=1,\dots,K$. The estimator could be iteratively updated.
However, if initial values are not appropriately chosen,
some elements of $\widehat{\ddelta}^{(i)}$ at the $i$th iteration are possible to jump out of the support of the threshold covariate $X_t$ or be very close to another kink point to make them hard to distinguish. We define the inadmissible set $\mathcal{T}$ as
$$
\mathcal{T}=\left\{\widehat{\delta}_k:  (\text{$\widehat{\delta}_k$ jumps out of the support set $\Omega$}) \bigcup(\text{$\widehat{\delta}_k$ is very close to another $\widehat{\delta}_{k^{'}}$})\right\}.
$$
The underlying reason is that the  local linear approximation  technique is sensitive to the initial values ${\ddelta}^{(0)}$, which makes the algorithm easy to get stuck in local optima. To deal with this drawback, we
iteratively update the initial values using the bootstrap samples to make the new algorithm insensitive to the original initial values. It shares the similar spirit of
the bootstrap restarting idea of \cite{wood2001minimizing}. 
Thus, we call it Bootstrap Restarting Iterative Segmented Quantile (BRISQ) regression algorithm.

\begin{figure}[!h]
\centering
\includegraphics[scale=0.5]{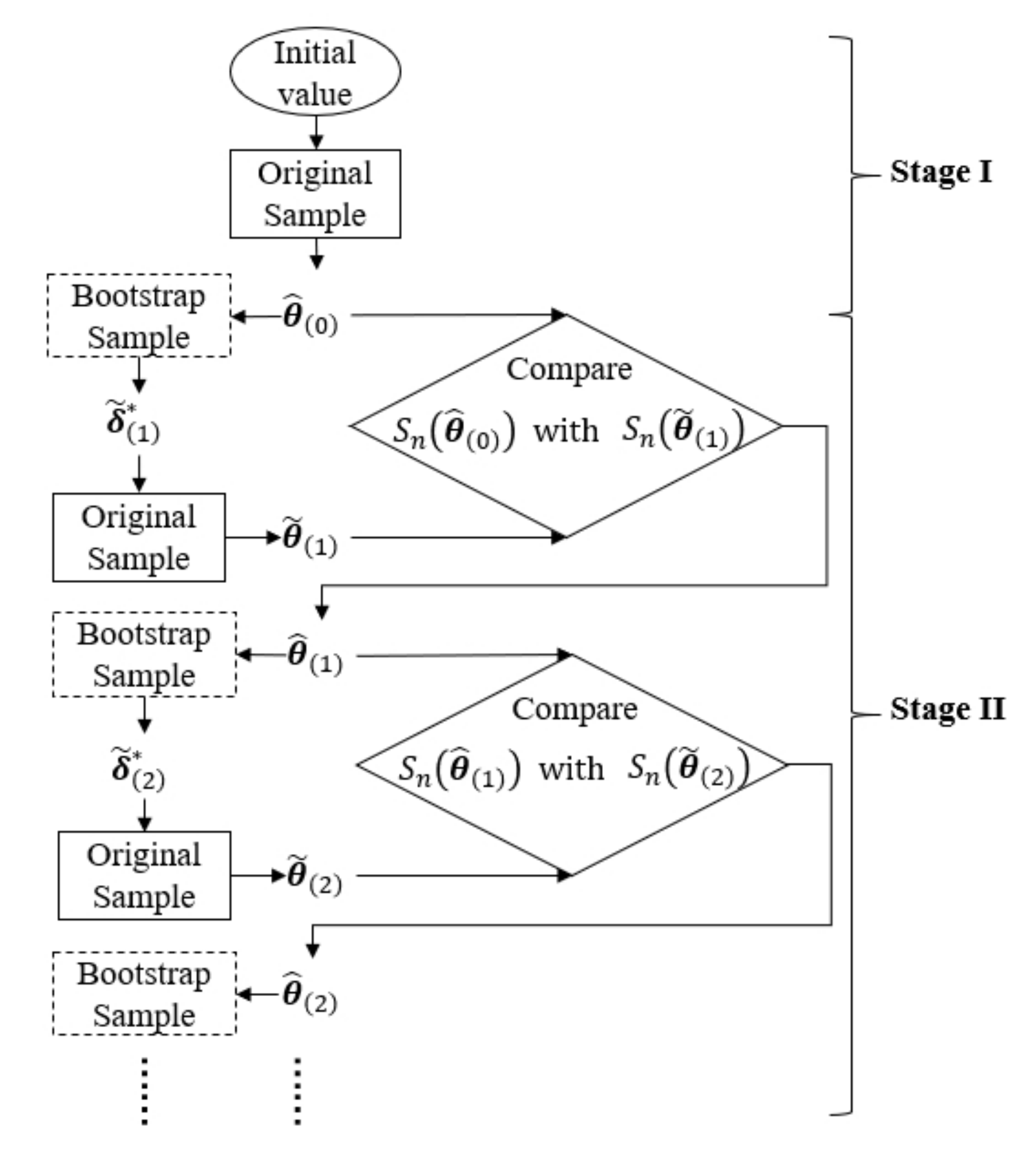}
\vspace{-4pt}
\caption[]{The flowchart of the BRISQ algorithm. Stage \uppercase\expandafter{\romannumeral1} contains Steps 1-3  and Stage \uppercase\expandafter{\romannumeral2} contains Step 4 in the Algorithm \ref{alg:1}.}
\label{fig:map}
\end{figure}

The main idea of the BRISQ algorithm is illustrated as follows. We first initialize parameters ${\ddelta}^{(0)}$ evenly dispersed on the domain of $X_t$ given $K$ and obtain the estimator $\widehat{\ttheta}_{(0)}=(\widehat{\eeta}_{(0)}\trans,\widehat{\ddelta}_{(0)}\trans)\trans$
by iteratively fitting the working model (\ref{eq:6}).
Then, we generate a bootstrap sample $\mathcal{X}_n^*$ in the classic way that we randomly select original observations with replacement and estimate Model (\ref{eq:6}) again using $\widehat{\ddelta}_{(0)}$ as the initial kink locations to obtain the bootstrap estimator
$\widetilde{\ttheta}^*_{(1)}=(\widetilde{\eeta}^{*\mbox{\tiny{T}}}_{(1)},
\widetilde{\ddelta}^{*\mbox{\tiny{T}}}_{(1)})\trans$.
And then, we estimate Model (\ref{eq:6}) again based on the original sample using the bootstrap estimator $\widetilde{\ddelta}^{*}_{(1)}$ as initial values and obtain the new estimator $\widetilde{\ttheta}_{(1)}=(\widetilde{\eeta}^{\mbox{\tiny{T}}}_{(1)},
\widetilde{\ddelta}^{\mbox{\tiny{T}}}_{(1)})\trans$.
Next, we compare $S_n(\widetilde{\ttheta}_{(1)})$ with $ S_n(\widehat{\ttheta}_{(0)})$.
If $S_n(\widetilde{\ttheta}_{(1)}) < S_n(\widehat{\ttheta}_{(0)})$, we update  $\widehat{\ttheta}_{(1)}=\widetilde{\ttheta}_{(1)}$; otherwise, $\widehat{\ttheta}_{(1)}=\widehat{\ttheta}_{(0)}$.
Last, we repeat the previous procedure until convergence.
The flowchart of this BRISQ algorithm is displayed in Figure \ref{fig:map}.
In practice, this algorithm can efficiently jump out of local minima and substantially improve the stability and accuracy of estimation, therefore less sensitive to the original initial values.
The detailed procedures are summarized in Algorithm  \ref{alg:1} and easily implemented using the newly developed R package \emph{MultiKink}.

\begin{algorithm}
\caption{Bootstrap Restarting Iterative Segmented Quantile (BRISQ) Algorithm.}\label{alg:1}
\begin{algorithmic}
\STATE {\bf Step 1.} Initialize parameters ${\ddelta}^{(0)}$ evenly dispersed on the domain of $X_t$; 
\STATE {\bf Step 2.} Fit Model (\ref{eq:6}) using the standard linear quantile regression based on the initial kink locations  ${\ddelta}^{(0)}$ and obtain estimators $\widehat{\bbeta}^{(1)}$ and $\widehat{\pphi}^{(1)}$ for $\bbeta$ and $\pphi$, respectively.
Update the kink locations estimators $\widehat{\ddelta}^{(1)}$ by $\widehat{\delta}_k^{(1)}={\delta}_k^{(0)}+\widehat{\phi}_k^{(1)}/\widehat{\beta}_k^{(1)}$, for $k=1,\dots,K$.
\STATE {\bf Step 3.} Repeat Step 2 iteratively  until convergence or any $\widehat{\delta}_k^{(i)}$ at the $i$th step falls into $\mathcal{T}$. The resulting estimator is denoted by $\widehat{\ttheta}_{(0)}=(\widehat{\eeta}_{(0)}\trans,\widehat{\ddelta}_{(0)}\trans)\trans$;
\STATE {\bf Step 4.}\quad \For{b=1:B}{
\STATE {\bf Step 4.1.} Generate a bootstrap sample $\mathcal{X}_n^*$.
\STATE {\bf Step 4.2.} Find $\widetilde{\ttheta}^*_{(b)}=(\widetilde{\eeta}^{*\mbox{\tiny{T}}}_{(b)},
\widetilde{\ddelta}^{*\mbox{\tiny{T}}}_{(b)})\trans$ for the bootstrap sample $\mathcal{X}_n^*$ using $\widehat{\ddelta}_{(b-1)}$ as the initial kink locations using the same procedures as Steps 2-3;
\STATE {\bf Step 4.3.}  Find  $\widetilde{\ttheta}_{(b)}=(\widetilde{\eeta}^{\mbox{\tiny{T}}}_{(b)},
\widetilde{\ddelta}^{\mbox{\tiny{T}}}_{(b)})\trans$  for  the original sample $\mathcal{X}_n$  using $\widetilde{\ddelta}^*_{(b)}$ as the initial kink locations using the same procedures as Steps 2-3;
\STATE {\bf Step 4.4.} Compare $S_n(\widetilde{\ttheta}_{(b)})$ with $ S_n(\widehat{\ttheta}_{(b-1)})$. If $S_n(\widetilde{\ttheta}_{(b)})<S_n(\widehat{\ttheta}_{(b-1)})$, $\widehat{\ttheta}_{(b)}=\widetilde{\ttheta}_{(b)}$; otherwise, $\widehat{\ttheta}_{(b)}=\widehat{\ttheta}_{(b-1)}$.
}
\STATE {\bf Step 5.} Obtain the final estimators $\widehat{\ttheta}=\frac{1}{B_\epsilon}\sum_{b_\epsilon}\widehat{\ttheta}_{(b_\epsilon)}$, where
$B_\epsilon=\#\left\{b_\epsilon:\Big|\frac{S_n(\widehat{\ttheta}_{(b_\epsilon)})-S_n(\widehat{\ttheta}_{\min})}{S_n(\widehat{\ttheta}_{\min})}\Big|\leq\epsilon\right\}$, $\widehat{\ttheta}_{\min}=\arg\min_{\{\widehat{\ttheta}_{(b)}: 1\le b\le B\}} S_n(\widehat{\ttheta}_{(b)})$.
\end{algorithmic}
\end{algorithm}

\subsubsection{Backward Elimination Algorithm for Estimating $K$} %
The aforementioned BRISQ algorithm works well when the exact number of kink points is given. However, the true number  denoted by  $K_0$ is usually unknown in practice. To estimate $K_0$, we first start with a large initial value $K_{\max}(\gg K_0)$ and then iteratively fit the working model (\ref{eq:6}) using the BRISQ algorithm in which we discard all $\widehat{\delta}_k$'s and corresponding $\widetilde{U}_k$'s and $\widetilde{V}_k$'s if $\widehat{\delta}_k\in\mathcal{T}$ at each iteration. When it stops, we find an estimator for the number of kink points, denoted by $K_*(<K_{\max})$.

However, we find that $K_*$  often overestimates the true value $K_0$ in practice.  To improve the selection and estimation accuracy, one can evaluate each MKQR model with $k=0,1,\cdots,K_*$ kink points according to a prescribed information criterion and find the final model with the smallest information criterion. In our algorithm, we suggest a strengthened quantile Bayesian information criterion (sBIC) to refine the kink points detection,
\begin{equation}\label{eq:7}
\text{sBIC}(K)=\log\left(S_n(\widehat{\ttheta}_K)\right)+N_{K}\frac{\log n}{2n}C_n,
\end{equation}
where $\widehat{\ttheta}_K$ denotes the estimator of parameters $\ttheta=(\eeta\trans,\ddelta\trans)\trans$ with $K$ kink points, $N_{K}$ equals to $2+p+2K$ and $C_n$ is a positive constant that allows to approach the infinity as $n$ increases.
When $C_n=1$, sBIC in (\ref{eq:7}) becomes the standard quantile BIC studied by \cite{lian2012note} for consistent model selection.  When $C_n>1$, it is similar to the modified BIC of \cite{lee2014model}.
The selection consistency of the sBIC selector will be demonstrated in the next section.
The BIC-type  criteria have been widely used in model selection. For example,  \cite{wang2007bic} proved that the BIC tuning parameter selector is  able to identify the true linear model consistently. \cite{chen2008ebic} further proposed an extended BIC (EBIC) to take into account both the model complexity and the sample size for consistent model selection. \cite{lee2014model} showed that a modified BIC is consistent in model selection for  high dimensional linear quantile regression.

To further improve the computational efficiency, we employ the backward elimination procedure for estimating the true value $K_0$. Given $K_*$, we re-estimate the new  MKQR model with $K_*-1$ kink points  using the BRISQ algorithm, and then compare the sBIC values of two models. This procedure is repeated until the sBIC values does not decrease. Then, the final  estimators for $K_0$ and  $\ttheta$ are obtained corresponding to the minimum sBIC. The detailed algorithm is summarized in Algorithm \ref{alg:2}. Simulations will illustrate that this algorithm is able to identify the true kink points consistently.

\begin{algorithm}
\caption{~~Backward Elimination  Algorithm for Estimating $K$. }\label{alg:2}
\begin{algorithmic}
\STATE {\bf Step 1.}
Given $K_{\text{max}}$ initial kink points, repeat Step 2 of Algorithm \ref{alg:1} iteratively and remove any $\widehat{\delta}_k$ at each iteration if $\widehat{\delta}_k\in\mathcal{T}$ until convergence. $K_*$ denotes the resulting estimated number of kink points.
\STATE {\bf Step 2.} Estimate the working model (\ref{eq:6}) with $K_*-1$ initial kink points using  Algorithm \ref{alg:1} to obtain $\widehat{\ttheta}_{K_*-1}$ and $\text{sBIC}(K_*-1)$.
\STATE {\bf Step 3.} If $\text{sBIC}(K_*-1)<\text{sBIC}(K_*)$, then update $K_*=K_*-1$ and go to {Step 2}; If $\text{sBIC}(K_*-1)\ge \text{sBIC}(K_*)$,
then stop, set $\widehat{K}=K_*$ and $\widehat{\ttheta}=\widehat{\ttheta}_{K_*}$.
If $K_*=0$, then stop and there is no kink point.
\end{algorithmic}
\end{algorithm}

\subsection{Asymptotic Properties}
\subsubsection{Selection Consistency}
To show the selection consistency of $\widehat{K}$, we need to introduce some notations. Denote  $h(\bW_t;\ttheta)=(1,X_t,(X_t-\delta_1)_+,\cdots,(X_t-\delta_{K})_+,\bZ_t\trans,-\beta_1I(X_t>\delta_1),\cdots,-\beta_KI(X_t>\delta_{K}))\trans$  and  the $\tau$th conditional quantiles of $e_t$ given $\bW_t$ as $F_t^{-1}(\tau|\bW_t)=\inf\{u:F(u|\bW_t)\geq0\}$. We then make the following assumptions.
\begin{itemize}
 \item[(A1)]  $F_t\equiv F(\cdot|\bW_t)$ has a continuous density $f_t(\cdot|\bW_t)$ that satisfies $0<\inf_tf_t(\cdot)<\sup_tf_t(\cdot)<\infty$ at the point $F^{-1}(\tau|\bW_t)$ for any sequence of values of $\bW_t$.
\item[(A2)] The matrix $E\left\{h(\bW_t;\ttheta)h\trans(\bW_t;\ttheta)\right\}$ is finite and positive definite.
\item[(A3)]  $C_n\log(n)/\sqrt{n}\rightarrow0$ as $n\rightarrow\infty$.
\end{itemize}

Assumption (A1) is  generally assumed in quantile regression. Assumption (A2) is similar to Assumption (A) in  \cite{lian2012note}.  Assumption (A3)  requires that $C_n=o(\sqrt{n}/\log(n))$ which means $C_n$ cannot
diverge too fast to the infinity as $n$ increases to avoid underfitting the true model.

\begin{thm}\label{thm1}
Under Assumptions (A1)-(A3), let $\widehat{K}=\arg\min_{k=0,\cdots,K_*}\text{sBIC}(k)$, we have $P(\widehat{K}=K_0)\rightarrow1$ as $n\rightarrow\infty$.
\end{thm}

 Theorem \ref{thm1} shows that the quantile sBIC is able to consistently select the true number of kink points.  This result plays a fundamental role in statistical inference since we will study the limiting distribution of the parameter estimators $\widehat{\ttheta}$ given the true number of kink points.

\subsubsection{Limiting Distribution}
Next, we derive the asymptotic properties for $\widehat{\ttheta}$.  Denote the true parameters as $\ttheta_0=(\eeta\trans_0,\ddelta\trans_0)\trans=\arg\min S(\ttheta)$, where $S(\ttheta)=E\left[\rho_\tau\{Y_t-Q_Y(\tau;\ttheta|\bW_t)\}\right]$. Define
\begin{align*}
\C_n&=E\left\{\left(\frac{\partial S_n(\ttheta)}{\partial\ttheta}\right)\left(\frac{\partial S_n(\ttheta)}{\partial\ttheta}\right)\trans\right\}\Big|_{\ttheta=\ttheta_0}
=n^{-1}\sum_{t=1}^n\tau(1-\tau)E\{h(\bW_t;\ttheta_0)h\trans(\bW_t;\ttheta_0)\},\\
\D_n&=E\left\{\frac{\partial^2S_n(\ttheta)}{\partial\ttheta\partial\ttheta\trans}\right\}\Big|_{\ttheta=\ttheta_0}
=n^{-1}\sum_{t=1}^n\frac{\partial}{\partial\ttheta}E[\psi_\tau\{Y_t-Q_{Y}(\tau;\ttheta|\bW_t)\}h(\bW_t;\ttheta)]\Big|_{{\ttheta}=\ttheta_0},
\end{align*}
where  $\psi_\tau(u)=\tau-I(u\leq0)$.
To establish the asymptotic distribution of $\widehat{\ttheta}$, we make the following assumptions.
\begin{itemize}
\item[(A4)] The objective function $S(\ttheta)$ has a unique global minimum at $\ttheta_0$.
\item[(A5)] The threshold variable $X_t$ has a continuous density function with a compact support $[-M,M]$, where $M$ is a positive constant.
\item[(A6)] $\max_{1\leq t\leq n}\parallel\Z_t\parallel=o_p(n^{1/2})$ and $E(\parallel\bZ\parallel^3)$ is bounded.
\item[(A7)] Given $K$ and $\bbeta\neq\bf{0}$, there exist a nonnegative definite matrix $\C$ and a full rank matrix $\D$, such that $\lim_{n\rightarrow\infty}\C_n=\C$ and $\lim_{n\rightarrow\infty}\D_n=\D$.
\end{itemize}

Assumption (A4) ensures the identifiability of estimation. Assumptions (A5)-(A6) impose some conditions on the threshold variable and other covariates, respectively, which can also be found in \cite{li2011bent} and \cite{zhang2017composite}.
Assumptions (A1) and (A4)-(A6) are used for the proof of consistency of $\widehat{\ttheta}$ and additional Assumption (A7) suffices for the asymptotical normality.
The following theorem demonstrates the limiting distribution of the proposed estimator for $\ttheta_0$.
\begin{thm}\label{thm2}
Suppose the true number $K$  of kink points in Model (\ref{eq:mkqr}) is given and  Assumptions (A1) and (A4)-(A7) hold, as $n\rightarrow\infty$, we have
$$
\sqrt{n}(\widehat{\ttheta}-\ttheta_0)\stackrel{d}{\longrightarrow}N(0,\SSigma),
$$
where  $\SSigma=\D^{-1}\C\D^{-1}$.
\end{thm}

According to Theorem \ref{thm2},  the regression coefficients $\eeta$ and the threshold parameters $\ddelta$ are jointly asymptotically normal with $\sqrt{n}$ convergence rate. In  conventional jump threshold model, the threshold parameter estimators converge to a nonstandard asymptotic distribution with $n$ convergence rate, see \cite{hidalgo2019robust} for more details. 

Moreover, we estimate $\SSigma$ by a plugging estimator   $\widehat{\SSigma}_n=\widehat{\D}_n^{-1}\widehat{\C}_n\widehat{\D}_n^{-1}$, where $\widehat{\C}_n=n^{-1}\sum_{t=1}^n\tau(1-\tau)h(\bW_t;\widehat{\ttheta})h\trans(\bW_t;\widehat{\ttheta})$ and $\widehat{\D}_n=n^{-1}\sum_{t=1}^n\hat{f}_t(\hat{e}_t)h(\bW_t;\widehat{\ttheta})h\trans(\bW_t;\widehat{\ttheta})$.  $\widehat{\D}_n$   requires  consistent estimate for conditional density function $f_t(\cdot)$ of error term $e_t$. We suggest using the method called \emph{Hendricks-Koenker Sandwich} based on the difference quotients discussed by \cite{hendricks1992hierarchical}. To select the bandwidth, two choices are often used. One is based on Edgeworth expansions of studentized quantiles described by \cite{hall1988distribution}, the other is based on the minimum of the mean squared error of the density estimator suggested by  \cite{bofingeb1975estimation}.
In our R package \emph{MultiKink}, we provide both versions to estimate the covariance matrix.

\csection{STATISTICAL INFERENCE}
\vspace{-0.5cm}
\subsection{Testing the Existence of Kink Effects}\label{sub3.1}
The kink effect estimation is meaningful if and only if the kink effect truly exists. In this section, we are interested in testing the  existence of kink effects in the conditional quantiles.  For $\tau\in (0,1)$, we consider the following null ($H_0$) and alternative ($H_1$)  hypotheses for the quantile regression (\ref{eq:mkqr}),
\begin{eqnarray} \label{test}
H_0: {\beta}_k={0}, \text{~for all~} k=1,\cdots,K.
\ \ \text{v.s.}\ \ \
H_1: {\beta}_k\ne {0}, \text{~for some~} k=1,\cdots,K.
\end{eqnarray}
Note that the parameters $\beta_k$'s depend on $\tau$. Under the null hypothesis, the MKQR model (\ref{eq:mkqr}) degenerates to an ordinary quantile regression without any kink point.
Under the alternative hypothesis, there exists at least one statistically significant kink point at the $\tau$th quantile. Thus, we suggest the following score-based test statistic
based on kink quantile regression with an unknown threshold,
\begin{equation}\label{eq:11}
T_n(\tau)=\sup_{\delta\in\Gamma}|R_n(\delta)|,
\end{equation}
where $R_n(\delta)=n^{-1/2}\sum_{t=1}^n\psi_\tau(Y_t-\widehat{\aalpha}\trans\V_t)(X_t-\delta)I(X_t\leq\delta),$
$\psi_\tau(u)=\tau-I(u\leq0)$, $\delta$ denotes the location of an unknown threshold, $\V_t=(1,X_t,\bZ\trans_t)\trans$, ${\aalpha}=(\alpha_0,\alpha_1,\ggamma\trans)\trans$ and $\widehat{\aalpha}=\arg\min_{\aalpha}\sum_{t=1}^n\rho_\tau(Y_t-\aalpha\trans\V_t)$. Note that $R_n(\delta)$ is essentially  the partial subgradient  of the objective function with respect to $\beta_1$ evaluated at $\beta_1=0$ and ${\aalpha}=\widehat{\aalpha}$ up to a constant in the model (\ref{eq:mkqr}) with $K=1$
\footnote{In fact, $R_n(\delta)$ is the partial subgradient  of the quantile objective function with respect to $\beta_1$ evaluated at $\beta_1=0$ and ${\aalpha}=\widehat{\aalpha}$ up to a constant for the model
$Q_Y(\tau;\ttheta|\bW_t)=\alpha_0+\alpha_1X_t+\beta_1(X_t-\delta_k)I(X_t\leq \delta)+\ggamma\trans\bZ_t,$
which is essentially same as the model (\ref{eq:mkqr}) with $K=1$ after simple reparameterizations.}.
$T_n(\tau)$  can be viewed as a weighted CUSUM test statistic based on the signs of quantile residuals.
Intuitively, under the null hypothesis, residuals $Y_t-\widehat{\aalpha}\trans\V_t$ are evenly located below or above zero which result in a relatively small value of $T_n(\tau)$.
On the other hand, under the alternative hypothesis, the model is misspecified and the residuals would be consistently positive or negative which implies the large values of $T_n(\tau)$.
The idea of subgradient-based tests has been studied in the literature. For example,  \cite{qu2008testing} constructed the subgradient test statistic in quantile regression for testing the structural changes.
\cite{zhang2014testing} proposed a score test based on the subgradient to test for the jumping threshold effect in threshold models.  \cite{zhang2017continuous} developed a related test for the continuous threshold effect in asymmetric least square regression.

Theorem \ref{thm3} in the following derives the asymptotic behavior of $T_n(\tau)$. We introduce some notations.
 Define $\bH_{1n}(\delta)=n^{-1}\sum_{t=1}^nE\left\{\V_t(X_t-\delta)I(X_t<\delta)f_t(e_t)\right\}$, $\bH_n=n^{-1}\sum_{t=1}^nE\left\{\V_t\V_t\trans f_t(e_t)\right\}$ and $\bH_{2n}(\delta,\beta_1)=n^{-1}\sum_{t=1}^nE\{\V_t\beta_1(X_t-\delta)I(X_t>\delta)f_t(e_t)\}$. We further   assume that
\begin{itemize}
\item[(A8)]  $\lim_{n\rightarrow\infty}\bH_n=\bH$, where $\bH$ is a positive definite matrix, and $\lim_{n\rightarrow\infty}\bH_{1n}=\bH_1$ and $\lim_{n\rightarrow\infty}\bH_{2n}=\bH_{2}$.
\item[(A9)] The  density function of $e_t$, $f_t(\cdot)$ for $t=1,\cdots,n$,  has a bounded first-order derivative.
\end{itemize}
\begin{thm}\label{thm3}
Suppose Assumptions (A1) and (A8)-(A9) hold, we have
\begin{equation}
T_n(\tau)\Rightarrow\sup_{\delta}|R(\delta)+q(\delta,\beta_1)|,
\end{equation}
where ``$\Rightarrow$'' denotes weak convergence, $R(\delta)$ is a Gaussian process with mean zeros and covariance function $W(\delta,\delta^{'})=\tau(1-\tau)E[\{(X_t-\delta)I(X_t\leq\delta)-\bH_1\trans(\delta)\bH^{-1}\V_t\}\{
(X_t-\delta^{'})I(X_t\leq\delta^{'})-\bH_1\trans(\delta^{'})\bH^{-1}\V_t\}]$ and $q(\delta,\beta_1)=-\bH_1(\delta)\bH^{-1}\bH_2(\delta,\beta_1)$.
\end{thm}

According to Theorem \ref{thm3}, under the null hypothesis, $q(\delta,\beta_1)=0$ and $R_n(\delta)$ would converge to a Gaussian process $R(\delta)$ with mean zeros. When the kink effect exists under the alternatives, $q(\delta,\beta_1)\neq0$ and $T_n(\tau)$ would be significantly larger than zero.
Therefore, large values of $T_n(\tau)$ provide the evidence against the null hypothesis. The  score-type statistic is only built on the null hypothesis without fitting models under the alternative hypothesis, so it  can also be directly used to test the existence of multiple kink points.
Since the asymptotical null distribution of $T_n(\tau)$ is nonstandard, we approximate the P-values using wild bootstrap \citep{feng2011wild}. The detailed procedures are relegated to Algorithm \ref{alg:3} in the Appendix.

\subsection{Confidence Intervals for Kink Parameters}
Next, we provides three types of confidence intervals (CI) for kink location parameters.
First, the traditional Wald-type $(1-\alpha)$th CIs are constructed based on the asymptotical normality in Theorem \ref{thm2}, i.e., $\widehat{\delta}_k\pm z_{\alpha/2}\mbox{SE}(\widehat{\delta}_k)$, for $k=1,\cdots,\widehat{K}$, where $z_{\alpha/2}$ is the $\alpha/2$ upper tailed critical value of the standard normal distribution and $\mbox{SE}(\widehat{\delta}_k)$ is the estimated standard error of $\widehat{\delta}_k$.
The Wald-type intervals involve estimation of the covariance matrix.
Second, the Bootstrap CIs are defined as $[\widehat{\delta}_{k,\alpha/2}^*,\widehat{\delta}_{k,1-\alpha/2}^*]$,
the $(\alpha/2)$th and $(1-\alpha/2)$th quantiles of bootstrap estimators
$\{\widehat{\delta}_{k,b}^*, b=1,2,\cdots, B\}$ with $B$ paired bootstrap samples.

Third, we construct a test-inversion confidence interval for $\ddelta$ based on a smoothed rank score test. Consider the following hypotheses for a given $\widetilde{\ddelta}$ in the domain of $X_t$ and $\tau\in(0,1)$, 
\begin{equation}\label{eq:13}
H_0: \ddelta=\widetilde{\ddelta}\quad\text{v.s.}\quad H_1: {\ddelta}\neq\widetilde{{\ddelta}.}
\end{equation}
Under $H_0$, we can obtain the estimator $\widehat{\eeta}(\widetilde{{\ddelta}})$ of the regression coefficients $\eeta$ given  $\widetilde{{\ddelta}}$ by fitting the standard linear quantile regression. \cite{Muggeo2017Interval} pointed that naive score statistic in threshold models may lower the test power due to the non-differentiable and non-smooth nature.
To deal with the non-smoothness of the indicator function $I(X_t>\delta_k)$,
we use the smoothed Gaussian distribution function $\Phi((X_t-{\delta}_k)/h_k)$  to approximate $I(X_t>\delta_k)$, where $h_k$ is the bandwidth. The smoothed objective function becomes
$$\widetilde{Q}_Y(\tau;\eeta,\ddelta|\bW_t)
=\alpha_0+\alpha_1X_t+\sum_{k=1}^K\beta_k(X_t-\delta_k)\Phi((X_t-{\delta}_k)/h_k)+\ggamma\trans\bZ_t.
$$
Then, we take the first partial derivative of $\widetilde{Q}_Y(\tau;\eeta,\ddelta|\bW_t)$ with respect to
$\ddelta$ evaluated at $\ddelta=\widetilde{\ddelta}$ and $\eeta=\widehat{\eeta}(\widetilde{{\ddelta}})$, denoted by
$\bP_t(\tau;\widehat{\eeta}(\widetilde{{\ddelta}}),\widetilde{{\ddelta}})
=(p_{t1},\cdots,p_{tK})\trans$, where $p_{tk}=-\widehat{\beta}_k\Phi((X_t-\widetilde{\delta}_k)/h_k)
-\widehat{\beta}_k(X_t-\widetilde{\delta}_k)\phi((X_t-\widetilde{\delta}_k)/h_k)h_k^{-1}$ and $\phi(\cdot)$ is the first derivative of $\Phi(\cdot)$.
 Motivated by the rank score tests
in \cite{Gutenbrunner1992rankscore}, \cite{Gutenbrunner1993test} and \cite{zhang2014testing},
we define a smoothed rank score (SRS) test statistic as
\begin{equation}\label{eq:14}
\mbox{SRS}_n(\tau)=\bS_n^{*\mbox{\tiny{T}}}\V_n^{-1}\bS_n^{*},
\end{equation}
where ${\bS}^*_n=n^{-1/2}\sum_{t=1}^n\bP_t^*\{\tau;\widehat{\eeta}(\widetilde{{\ddelta}}),\widetilde{{\ddelta}}\}\psi_\tau(\tilde{e}_t)$, ${\tilde{e}}_t=Y_t-Q_Y\{\tau;\widehat{\eeta}(\widetilde{{\ddelta}}),\widetilde{{\ddelta}}|\bW_t\}$ is the $t$th residual under $H_0$, and ${\V}_n=n^{-1}\sum_{t=1}^n\tau(1-\tau)\bP_t^*\{\tau;\widehat{\eeta}(\widetilde{{\ddelta}}),\widetilde{{\ddelta}}\}\bP_t^*\{\tau;\widehat{\eeta}(\widetilde{{\ddelta}}),\widetilde{{\ddelta}}\}\trans$. Here, $\bP_t^*\{\tau;\widehat{\eeta}(\widetilde{{\ddelta}}),\widetilde{{\ddelta}}\}$ is defined as follows. Let
$\M_t(\widetilde{\ddelta})=(1,X_t,(X_t-\widetilde{\delta}_1)_+,\cdots,(X_t-\widetilde{\delta}_K)_+,\bZ_t\trans)\trans$
and $\M(\widetilde{\ddelta})=(\M_1(\widetilde{\ddelta}),\cdots,\M_n(\widetilde{\ddelta}))\trans$. Let $\bP=(\bP_1\{\tau;\widehat{\eeta}(\widetilde{{\ddelta}}),\widetilde{{\ddelta}}\},\cdots,
\bP_n\{\tau;\widehat{\eeta}(\widetilde{{\ddelta}}),\widetilde{{\ddelta}}\})\trans$,
and define  $\bP^*=(\I_n-\LLambda)\bP$, where $\I_n$ is an $n\times n$ identity matrix, $\LLambda=\M(\widetilde{{\ddelta}})\{\M\trans(\widetilde{{\ddelta}})\PPsi\M(\widetilde{{\ddelta}})\}^{-1}\M\trans(\widetilde{{\ddelta}})\PPsi$, $\PPsi=\mbox{diag}(\hat{f}_1(\tilde{e}_1),\cdots,\hat{f}_n(\tilde{e}_n))$.
$\bP_t^*\{\tau;\widehat{\eeta}(\widetilde{{\ddelta}}),\widetilde{{\ddelta}}\}$
is defined as the $t$th row of the $n\times K$ matrix $\bP^*$ which is considered as the residuals by projecting the partial score vector $\bP$ on  $\M(\widetilde{\ddelta})$.

Intuitively, under $H_0$, the partial scores tend to be zero which implies that the test statistic is relatively small; otherwise, the large test statistic values provide the strong evidence against $H_0$. The following proposition demonstrates the null asymptotic distribution of $\mbox{SRS}_n(\tau)$.

\begin{prop}\label{prop1}
Suppose that Assumptions (A1) and (A6)-(A9) hold. Under the null hypothesis $H_0$ in (\ref{eq:13}) for any $\tau\in (0,1)$, as $n\rightarrow\infty$, we have $\mbox{SRS}_n(\tau)\stackrel{d}{\longrightarrow}\chi_{K}^2$.
\end{prop}

For one kink point $\delta$, the confidence interval can be obtained by inverting the rank score test due to the fact that the test statistic $\mbox{SRS}_n(\tau)$ is convex in $\delta$.  Specially, we first obtain the estimator $\widehat{\delta}$ for $\delta$ and then test $H_0: \delta=\widetilde{\delta}$ for $\widetilde{\delta}=\widehat{\delta}+\varrho$, where $\varrho$ is a small positive increment. If $H_0$ is not rejected, then increase $\widetilde{\delta}$ by $\widetilde{\delta}=\widetilde{\delta}+\varrho$ and  test $H_0: \delta=\widetilde{\delta}$ again. We repeat the previous testing procedure until $H_0$ is rejected and set the upper bound of the confident interval for $\delta$ as the minimum rejection point, denoted by $\widehat{\delta}_u$. In the similar way, we obtain the lower bound $\widehat{\delta}_l$. Thus, we can obtain a $(1-\alpha)$th confidence interval for $\delta$, $[\widehat{\delta}_l,\widehat{\delta}_u]$. For multiple kink model (\ref{eq:mkqr}), we separately construct the confidence interval for each kink  location parameter by controlling other kink estimators. The details are summarized in Algorithm \ref{alg:4} of the Appendix.


\csection{SIMULATION STUDIES}
\vspace{-0.5cm}
\subsection{Parameters Estimation}
We generate data from the following model
\begin{equation}
Y_t=\alpha_0+\alpha_1X_t+\sum_{k=1}^K\beta_k(X_t-\delta_k)I(X_t>\delta_k)+\gamma Z_t+\sigma(X_t,Z_t)e_t,\quad t=1,\dots,n,
\end{equation}
where $X_t\sim U(-5,5)$, $Z_t\sim N(1,1^2)$,  $e_t\sim N(0,1)$ or $t_3$ distribution and $\sigma(X_t,Z_t)$ controls the heteroscedasticity.  Specially, $\sigma(X_t,Z_t)$ equals to 1 for a homoscedastic model and $1+0.2X_t$ for a heteroscedastic model. We set $\alpha_0=1, \alpha_1=1,\gamma=1$
and consider three different cases for kink effects:
(1) $K=1$, $\beta_1=-3$ and $\delta_1=0.5$;
(2) $K=2$, $(\beta_1,\beta_2)=(-3,4)$ and $(\delta_1,\delta_2)=(-1,2)$; (3) $K=3$, $(\beta_1,\beta_2,\beta_3)=(-3,4,-4)$ and $(\delta_1,\delta_2,\delta_3)=(-3,0,3)$.


We first check the selection consistency of Theorem \ref{thm1}. Since Condition (A3) requires that $C_n\log(n)/\sqrt{n}\rightarrow0$ as $n\rightarrow\infty$, we consider $C_n=1,\log(\log(n))$ and $\log(n)$ in the definition of sBIC. We set the sample size $n=500$.
Table \ref{tab:1} reports the percentages of correctly selecting $\widehat{K}=K$ based on 1000 replications
under both homoscedastic and heteroscedastic models. All selection rates are very high and close to 100\%, especially when $C_n=\log(n)$. It shows that a diverging number for $C_n$ is favorable of  identifying the true model when the number of parameters is not fixed. It is in accord with \cite{fryzlewicz2014wild} which proposed a strengthened BIC for sequential change points detection.
We set $C_n=\log(n)$ for the rest of simulation studies.
These results validate the selection consistency of Theorem \ref{thm1}.

\begin{table}[ht]
\centering
\caption{The percentages of correctly selecting  $\widehat{K}=K$ for $C_n= 1, \log(\log(n)), \log(n)$.\label{tab:1}}
\begin{threeparttable}
\footnotesize
\begin{tabular}{ccccccccccc}
\hline
\hline
  \multirow{2}{*}{$e_t$} & \multirow{2}{*}{$\tau$} &\multicolumn{3}{c}{$K=1$} & \multicolumn{3}{c}{$K=2$} & \multicolumn{3}{c}{$K=3$}\\
\cmidrule(lr){3-5} \cmidrule(lr){6-8} \cmidrule(lr){9-11}
 & & 1 & $\log(\log(n))$ & $\log(n)$ & 1 & $\log(\log(n))$ & $\log(n)$ & 1 & $\log(\log(n))$ & $\log(n)$\\
\hline
\multicolumn{11}{c}{Homoscedasticity}\\
  $N$    & 0.3 & 91.6\% & 99.8\%  & 100.0\% & 92.8\% & 99.1\% & 99.6\% & 92.3\% & 97.5\% & 98.5\%\\
         &0.5  & 94.2\% & 100.0\% & 100.0\% & 92.4\% &99.6\%  & 99.6\% & 95.6\% & 99.0\% & 99.6\%\\
         &0.7  & 91.1\% & 98.8\%  & 100.0\% & 91.5\% & 99.6\% & 99.6\% & 94.8\% & 97.7\% & 98.9\% \\
     \cmidrule(lr){2-11}
 $t_3$  & 0.3 & 96.1\% & 99.8\% &  99.8\% & 95.5\% & 99.8\% & 99.6\% & 96.0\% & 98.3\% & 97.5\%\\
        & 0.5 & 98.0\% & 100.0\% & 100.0\% & 96.8\% & 100.0\% & 99.8\% & 97.8\% & 99.4\% & 99.0\%\\
        & 0.7 & 96.7\% & 99.8\% & 100.0\% & 95.3\% & 99.1\% & 99.8\%  & 94.8\% & 99.8\% & 96.6\%\\
\hline
\multicolumn{11}{c}{Heteroscedasticity}\\
  $N$   & 0.3 & 84.5\% & 97.3\% & 100.0\% & 86.8\% & 98.3\% & 99.8\% & 90.2\% & 97.7\% & 97.7\%\\
        &0.5  & 87.2\% & 96.2\% & 100.0\% & 87.0\% & 98.7\% & 99.8\% & 91.3\% & 98.1\% & 98.6\%\\
        &0.7 &  83.5\% & 97.1\% & 100.0\% & 83.8\% & 97.6\% & 99.8\% & 89.3\% & 97.1\% & 98.8\% \\
     \cmidrule(lr){2-11}
 $t_3$  & 0.3 & 89.5\% & 99.8\% & 99.8\% & 89.5\% & 98.0\% & 99.6\% & 95.4\% & 98.8\% & 99.2\%\\
        & 0.5 & 91.6\% & 98.4\% & 100.0\% & 92.1\% & 98.1\% & 99.8\% & 96.1\% & 98.8\% & 97.9\%\\
        & 0.7 & 91.4\% & 98.6\% & 100.0\% & 91.6\% & 98.1\% & 99.4\%  & 94.0\% & 99.8\% & 99.8\%\\
\hline
\end{tabular}
\end{threeparttable}
\end{table}

Next, we evaluate the finite sample performance of  parameter estimators to check the validity of Theorem \ref{thm2}. For Case (1) with single kink effect,  we compare the proposed estimation method with  the bent line quantile estimators  proposed by \cite{li2011bent} and the kink regression least squares  estimators proposed by \cite{hansen2017regression}. Both existing methods assume
there is only a single kink effect. We denote two methods as SKQR and SKLS, short for  Single Kink Quantile Regression  and Single Kink Least Square, respectively. We conduct the simulations 500 times and report the estimation biases (Bias), the empirical standard deviations (SD) and the mean square errors (MSE) for each parameter based on 500 estimates as well as their average estimated standard errors (SE) based on the asymptotical variance in Theorem \ref{thm2}. All simulation results are summarized in Table \ref{tab:2}. All estimates have ignorable biases, and the standard deviations (SD) are close to the estimated standard errors (SE). In the homoscedastic model with normal random errors, the SKLS estimators have smaller mean square errors (MSE) than others. However, when the model is heteroscedastic or the errors follow $t_3$ distribution, our method works better than the other two in terms of MSE, which demonstrate the robustness and efficiency of the proposed estimators.

\begin{table}[!htb]
\centering
\begin{threeparttable}
\footnotesize
\caption{Estimation results (multiplied by a factor of 10) of three methods for Case (1).}
\label{tab:2}
\begin{tabular}{ccccccccccccc}
\hline
\hline
\multirow{2}{*}{$e_t$} & & & \multicolumn{5}{c}{Homoscedasticity} & \multicolumn{5}{c}{Heteroscedasticity}\\
\cmidrule(lr){4-8} \cmidrule(lr){9-13} 
	&		&		&	$\alpha_0$	&	$\alpha_1$	&	$\beta_1$	&	$\gamma$	&	$\delta$	&	$\alpha_0$	&	$\alpha_1$	&	$\beta_1$	&	 $\gamma$	&	$\delta$	\\
\hline
$N$		&	SKQR	&	Bias	&	-0.020 	&	-0.014 	&	-0.002 	&	-0.058 	&	0.089 	&	0.054 	&	0.001 	&	-0.042 	&	0.022 	&	-0.002 	 \\
	&		&	SD	&	1.481 	&	0.480 	&	0.572 	&	0.795 	&	0.875 	&	1.081 	&	0.236 	&	0.246 	&	1.086 	&	0.888 	\\
	&		&	SE	&	1.415 	&	0.470 	&	0.576 	&	0.808 	&	0.752 	&	1.110 	&	0.272 	&	0.368 	&	1.018 	&	0.829 	\\
	&		&	MSE	&	0.218 	&	0.023 	&	0.033 	&	0.063 	&	0.077 	&	0.117 	&	0.006 	&	0.006 	&	0.117 	&	0.078 	\\
	&	SKLS	&	Bias	&	0.068 	&	-0.017 	&	-0.009 	&	-0.028 	&	0.019 	&	0.096 	&	-0.118 	&	-0.003 	&	-0.005 	&	0.036 	 \\
	&		&	SD	&	1.124 	&	0.444 	&	0.358 	&	0.612 	&	0.669 	&	1.074 	&	0.519 	&	0.261 	&	0.863 	&	0.683 	\\
	&		&	SE	&	1.132 	&	0.444 	&	0.377 	&	0.639 	&	0.599 	&	1.056 	&	0.508 	&	0.264 	&	0.849 	&	0.692 	\\
	&		&	MSE	&	\bf{0.126} 	&	\bf{0.020} 	&	\bf{0.013} 	&	\bf{0.037} 	&	\bf{0.045} 	&	0.116 	&	0.028 	&	0.007 	&	\bf{0.074} 	 &	\bf{0.047} 	\\
&	MKQR	&	Bias	&	0.051 	&	-0.010 	&	-0.031 	&	-0.052 	&	0.114 	&	0.077 	&	-0.043 	&	0.007 	&	0.011 	&	-0.014 	 \\
	&		&	SD	&	1.482 	&	0.577 	&	0.475 	&	0.779 	&	0.822 	&	1.036 	&	0.243 	&	0.226 	&	1.075 	&	0.843 	\\
	&		&	SE	&	1.435 	&	0.555 	&	0.480 	&	0.813 	&	0.758 	&	0.998 	&	0.234 	&	0.219 	&	1.022 	&	0.818 	\\
	&		&	MSE	&	0.219 	&	0.033 	&	0.023 	&	0.061 	&	0.069 	&	\bf{0.107} 	&	\bf{0.006} 	&	\bf{0.005} 	&	0.115 	&	0.071 	 \\
\hline
$t_3$	&	SKQR	&	Bias	&	0.109 	&	0.021 	&	-0.059 	&	0.095 	&	-0.101 	&	0.064 	&	0.017 	&	0.001 	&	-0.026 	&	-0.097 	 \\
	&		&	SD	&	1.539 	&	0.481 	&	0.707 	&	0.784 	&	0.909 	&	1.173 	&	0.268 	&	0.303 	&	1.165 	&	1.024 	\\
	&		&	SE	&	1.606 	&	0.543 	&	0.641 	&	0.923 	&	0.865 	&	1.304 	&	0.335 	&	0.478 	&	1.152 	&	0.936 	\\
	&		&	MSE	&	0.237 	&	\bf{0.023} 	&	0.050 	&	\bf{0.062} 	&	0.083 	&	0.137 	&	0.007 	&	0.009 	&	0.135 	&	0.105 	 \\
	&	SKLS	&	Bias	&	0.025 	&	-0.048 	&	0.019 	&	0.033 	&	-0.064 	&	-0.058 	&	-0.004 	&	-0.013 	&	0.104 	&	-0.083 	 \\
	&		&	SD	&	1.907 	&	0.766 	&	0.613 	&	1.037 	&	1.029 	&	1.887 	&	0.857 	&	0.481 	&	1.487 	&	1.322 	\\
	&		&	SE	&	1.893 	&	0.741 	&	0.628 	&	1.084 	&	1.010 	&	1.764 	&	0.852 	&	0.440 	&	1.449 	&	1.170 	\\
	&		&	MSE	&	0.362 	&	0.059 	&	0.037 	&	0.107 	&	0.106 	&	0.355 	&	0.073 	&	0.023 	&	0.221 	&	0.175 	\\
		&	MKQR	&	Bias	&	0.063 	&	-0.055 	&	0.009 	&	0.105 	&	-0.086 	&	0.087 	&	-0.001 	&	0.023 	&	-0.054 	&	-0.080 	 \\
	&		&	SD	&	1.495 	&	0.706 	&	0.474 	&	0.801 	&	0.891 	&	1.138 	&	0.307 	&	0.256 	&	1.181 	&	0.962 	\\
	&		&	SE	&	1.560 	&	0.607 	&	0.526 	&	0.892 	&	0.841 	&	1.080 	&	0.264 	&	0.239 	&	1.125 	&	0.889 	\\
	&		&	MSE	&	\bf{0.223} 	&	0.050 	&	\bf{0.022} 	&	0.065 	&	\bf{0.080} 	&	\bf{0.130} 	&	\bf{0.007} 	&	\bf{0.007} 	&	 \bf{0.135} 	&	\bf{0.093} 	\\
 \hline
\end{tabular}
\begin{tablenotes}
        \footnotesize
       \item  Bias: the empirical bias; SD: the empirical standard deviation; MSE: the mean square error; SE: the average estimated standard error. The minimum MSE among three estimators is highlighted in bold.
      \end{tablenotes}
\end{threeparttable}
\end{table}

For the multi-kink models, both SKLS and SKQR methods are not able to detect the multiple kink points. Thus, we only present the simulation results of the proposed MKQR estimators for Case (2) with $K=2$ in Table \ref{tab:case2}. All the biases are sufficiently close to zero
 and the SEs are compatible with the SDs for both homoscedastic and heteroscedastic errors.
To save the space, we omit the similar simulation results for Case (3) with $K=3$.
These results demonstrate the validity of Theorem \ref{thm2} for the multiple kink effects.


\begin{table}
\centering
\begin{threeparttable}
\footnotesize
\caption{Estimation results (multiplied by a factor of 10) of the proposed MKQR method for Case (2) with $K=2$.}
\label{tab:case2}
\begin{tabular}{cccccccccc}
\hline
\hline
  $e_t$ &$\tau$	&		&	$\alpha_0$	&	$\alpha_1$ &	$\gamma$		&	$\beta_1$	&	$\beta_2$	&	$\delta_1$	&	$\delta_2$	\\
\hline
\multirow{18}{*}{$N(0,1)$} & \multicolumn{9}{c}{Homoscedasticity}\\
 &0.2&	
 Bias	&	0.001   &	0.001  	&	0.001 		&	0.000 	&	0.000 	&	0.000 	&	0.001 	\\
 &&	SD	&	0.067 	&	0.020    &	0.014  	&	0.037 	&	0.044 	&	0.024 	&	0.019 	\\
&&	SE	&	0.067 	&	0.020    &	0.014 		&	0.038 	&	0.045 	&	0.024 	&	0.019 	\\
&&	MSE	&	0.005 	&	0.000    &	0.000  	&	0.001 	&	0.002 	&	0.001 	&	0.000 	\\
\cmidrule(lr){2-10}
&0.5	
&	Bias	&	-0.001 	&	0.000  	&	0.000  	&	-0.001 	&	0.002 	&	0.000 	&	-0.001 	\\
&	&	SD	&	0.061 	&	0.019   &	0.014 		&	0.035 	&	0.042 	&	0.022 	&	0.019 	\\
&	&	SE	&	0.065 	&	0.019   &	0.017 		&	0.043 	&	0.049 	&	0.026 	&	0.022 	\\
&	&	MSE	&	0.004 	&	0.000	  &	0.000  	&	0.001 	&	0.002 	&	0.000 	&	0.000 	\\
\cmidrule(lr){2-10}
&\multicolumn{9}{c}{Heteroscedasticity}\\
&0.2
&	Bias&	0.002   &	0.001 	    &	0.000 		&	0.001 	&	0.000 	&	-0.001 	&	0.001 	\\
&&	SD	&	0.032   &	0.007 	   &	0.007 		&	0.035 	&	0.063 	&	0.019 	&	0.027 	\\
&&	SE	&	0.032   &	0.007 	   &	0.006 		&	0.035 	&	0.064 	&	0.019 	&	0.027 	\\
&&	MSE	&	0.001	&	0.000 	   &	0.000  	&	0.001 	&	0.004 	&	0.000 	&	0.001 	\\
\cmidrule(lr){2-10}
&0.5
&	Bias&	-0.001 	&	0.000    	&	0.000  	&	0.000 	&	0.000 	&	-0.001 	&	-0.002 	\\
&&	SD	&	0.030 	&	0.007     &	0.006 		&	0.032 	&	0.059 	&	0.018 	&	0.026 	\\
&&	SE	&	0.030 	&	0.006     &	0.006 		&	0.033 	&	0.060 	&	0.018 	&	0.025 	\\
&&	MSE	&	0.001 	&	0.000	    &	0.000  	&	0.001 	&	0.003 	&	0.000 	&	0.001 	\\
\hline
&\multicolumn{9}{c}{Homoscedasticity}\\
\multirow{18}{*}{$t_3$}&0.2	
 &	Bias	&	0.007 	 	&	0.001      &	-0.002 	&	-0.017 	&	0.021 	&	0.005 	&	-0.003 	\\
&	&	SD	&	0.333 	 &	0.101 	     &	0.072 	&	0.175 	&	0.227 	&	0.132 	&	0.097 	\\
&	&	SE	&	0.318 	 &	0.097 	     &	0.067 	&	0.181 	&	0.216 	&	0.114 	&	0.093 	\\
&	&	MSE	&	0.111 		&	0.010       &	0.005 	&	0.031 	&	0.052 	&	0.017 	&	0.009 	\\
\cmidrule(lr){2-10}
&0.5	
&	Bias	&	-0.010  	&	-0.001    	&	0.002  	&	-0.007 	&	0.013 	&	0.004 	&	-0.003 	\\
&	&	SD	&	0.299 	 &	0.092 	       &	0.060 	&	0.161 	&	0.196 	&	0.119 	&	0.089 	\\
&	&	SE	&	0.278 	 &	0.085 	       &	0.060 	&	0.158 	&	0.188 	&	0.101 	&	0.081 	\\
&	&	MSE	&	0.089 		&	0.009         &	0.004 	&	0.026 	&	0.038 	&	0.014 	&	0.008 	\\
	\cmidrule(lr){2-10}
&\multicolumn{9}{c}{Heteroscedasticity}\\
&0.2	
&	Bias	&	0.009  	&	0.003   	&	-0.001 	&	-0.016 	&	0.035 	&	0.001 	&	0.002 	\\
&	&	SD	&	0.147  &	0.032 	   	&	0.033 	&	0.174 	&	0.320 	&	0.098 	&	0.142 	\\
&	&	SE	&	0.152 	&	0.033 	   &	0.030 	&	0.168 	&	0.308 	&	0.089 	&	0.130 	\\
&	&	MSE	&	0.022 	&	0.001 	   &	0.001 	&	0.030 	&	0.104 	&	0.010 	&	0.020 	\\
\cmidrule(lr){2-10}
&0.5	
&	Bias	&	-0.005 	&	-0.001	     &	0.000  	&	-0.005 	&	0.024 	&	0.000 	&	0.001 	\\
&	&	SD	&	0.135 	&	0.030 	     &	0.026 	&	0.155 	&	0.278 	&	0.087 	&	0.133 	\\
&	&	SE	&	0.131 	&	0.028 	     &	0.026 	&	0.148 	&	0.269 	&	0.080 	&	0.113 	\\
&	&	MSE	&	0.018 	&	0.001 	     &	0.001 	&	0.024 	&	0.078 	&	0.008 	&	0.018 	\\
 \hline
\end{tabular}
\begin{tablenotes}
        \footnotesize
        \item  Bias: the empirical bias; SD: the empirical standard deviation; MSE: the mean square error; SE: the average estimated standard error.
      \end{tablenotes}
\end{threeparttable}
\end{table}

\subsection{Power Analysis}
\begin{figure}[ht!]
\begin{center}
\includegraphics[scale=.95]{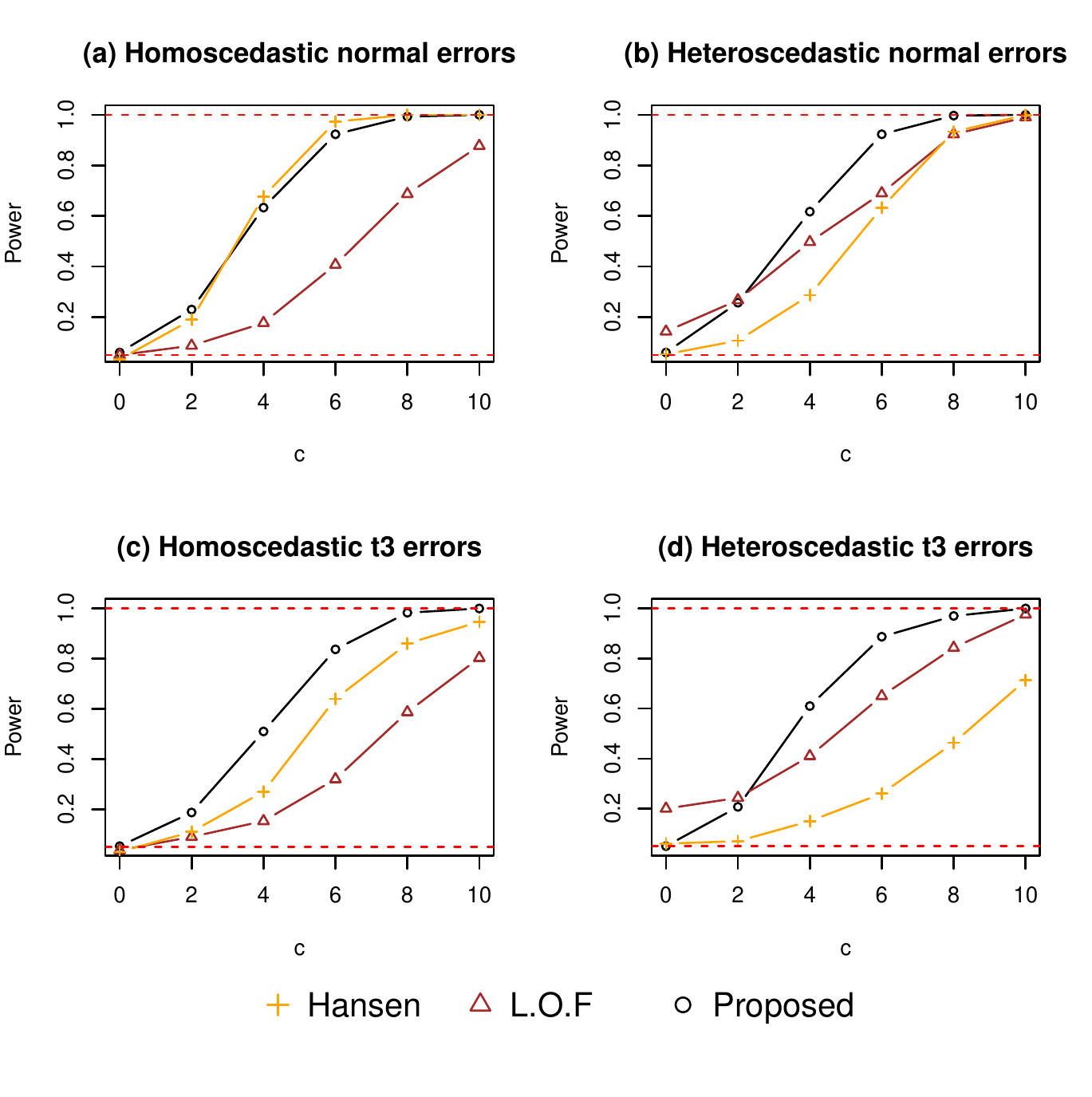}
\end{center}
\vspace{-1cm}
\caption{Power comparison of the proposed test at $\tau=0.5$ (black circle), the lack-of-fit test at $\tau=0.5$ (brown triangle) and the F-type test (orange plus) at the  significance level 5\%.}%
\label{fig:power}
\end{figure}

We now assess the power performances for testing the existence of kink effects in Section \ref{sub3.1}.
We generate the data from Case (1) except $\beta_1=n^{-1/2}c$, where $n=1000$, $c=0,2,4,6,8,10$ and $c=0$ corresponds to the null hypothesis.
We compare our proposed test with two existing tests,
the  lack-of-fit (L.O.F) test proposed by \cite{he2003lack} and the F-type test proposed by \cite{hansen2017regression}. The lack-of-fit test is a general test for checking model specification, which was also used in \cite{li2011bent}.
For our score-based test, we compute the P-values using wild bootstrap in Algorithm \ref{alg:3} with 300 replicates. Figure \ref{fig:power} displays the power curves of three tests over different signal strength values of $c$. Under the null hypothesis when $c=0$,
all methods have satisfactory  type I errors close to the nominal significance level 5\% for homoscedastic errors.
However, the L.O.F test can not control the type I errors when there exists heteroscedasticity. As $c$ increases, i.e. the kink effect gets enhanced, the empirical powers
to identify the kink effect for all methods gradually increase to one for each scenario.
Our proposed test have the higher empirical powers than the other two tests, especially when the errors follow the $t_3$ distribution or exist the heteroscedasticity.

\subsection{Confidence Intervals}
Last, we evaluate the finite sample performances of three confidence intervals for the kink location parameter $\ddelta$, the Wald-type CIs, the bootstrap CIs and the smoothed rank score (SRS) test-inversion CIs. We generate data from the MKQR model of Case (2) with $K=2$ and $n=500$.
To save the space, we only report the simulation results for the heteroscedastic model with the errors from $t_3$ distribution. Table \ref{tab:3} reports the coverage probabilities and  the mean width of 95\% confidence intervals as well as the average running time per replication at different quantile levels $\tau=0.3,0.5,0.8$ based on 1000 simulations.

\begin{table}
\centering
\begin{threeparttable}
\footnotesize
\caption{ 95\% confidence intervals for each kink point parameter in Case (2) in the heteroscedastic model with the $t_3$ errors.}
\label{tab:3}
\begin{tabular}{ccccccc}
\hline
\hline
 \multirow{2}{*}{$\tau$} & \multirow{2}{*}{Type} &  \multicolumn{2}{c}{Coverage probability} & \multicolumn{2}{c}{Mean interval length} & \multirow{2}{*}{Time(s)}\\
\cmidrule(lr){3-4} \cmidrule(lr){5-6}
  &  & $\delta_1$ & $\delta_2$  & $\delta_1$ & $\delta_2$  & \\
\hline
 0.3 & Wald  & 0.933 & 0.923  & 0.353 & 0.518 & 5.07 \\
          & Boot  & 0.964 & 0.970  & 0.391 & 0.689  & 377.81\\
          & Score   & 0.930 & 0.957  & 0.343 & 0.641 & 12.70\\
    0.5   & Wald  & 0.923 & 0.933  & 0.307 & 0.436 & 4.45  \\
         & Boot  & 0.968 & 0.982  & 0.323 &0.575  & 375.07 \\
          & Score   & 0.927 & 0.953  & 0.303 & 0.530 & 11.38\\
    0.8   & Wald  & 0.913 & 0.883  & 0.451 & 0.619 &4.53  \\
         & Boot  & 0.970 & 0.974  & 0.501 & 0.801 & 378.38\\
         & Score   & 0.917 & 0.930  & 0.449 & 0.774 & 13.82 \\
 \hline
\end{tabular}
\begin{tablenotes}
        \footnotesize
       \item  Wald: Wald-type CIs; Boot: bootstrap CIs; Score: SRS test-inversion CIs. Time is the average running time for one simulation.
      \end{tablenotes}
\end{threeparttable}
\end{table}

From  Table \ref{tab:3}, the coverage probabilities of Wald-type intervals are generally lower than the 95\% nominal level. \cite{hansen2017regression} and \cite{fong2017model} also found that Wald-type CIs have poor finite sample performance, especially for threshold parameters due to the parameter-effects curvature.
The bootstrap intervals have the highest coverage rates, but they have the largest interval lengths and need much more computing time. The bootstrap method is less computationally efficient.
The proposed smoothed rank score  (SRS) test-inversion CIs provide a balance between the estimation accuracy and the computation efficiency. They have higher coverage probabilities than the Wald-type intervals and also need much less computing time than the Bootstrap intervals.

\csection{EMPIRICAL ANALYSIS}
\vspace{-0.5cm}
\subsection{Secondary Industrial Structure of China}
The past few decades have witnessed the miracle of China's economic growth. Since China introduced the policy of reform and opening in 1978, GDP per capita has experienced a considerable growth and the industrial structure has also undergone tremendous changes.  Classical development economic theory  tells us that in the process of  development,  the proportion of first industry  decreases while the tertiary industry instead increases gradually for one country. Meanwhile, the proportion of secondary industry experiences a process of increasing rapidly at first and then gradually stops growing or even decreases, which implies the presence of a kink pattern. The economic development model of China, as the biggest developing country in the world, has been aroused a great of research interest, see \cite{song2011growing},  \cite{brandt2013factor},  \cite{cao2013agricultural} and etc.

In this section, we aim to investigate whether there exist kink effects between secondary industrial structure and the economic growth from the quantile regression perspective using the prefecture-level cities data in China. After removing the missing values, we collect data for 280 Chinese prefecture-level cities of year 2016 from the Organisation for Economic Co-operation and Development (OECD) database available at \url{https://insights.ceicdata.com/}. We consider the MKQR model
\begin{equation}\label{eq:15}
Q_Y(\tau|X_t,\bZ_t)=\alpha_0+\alpha_1X_t+\sum_{k=1}^K\beta_k(X_t-\delta_k)I(X_t>\delta_k)+\ggamma\trans\bZ_t,\quad t=1,\cdots,280,
\end{equation}
where $Y_t$ represents the proportion of secondary industry of the $t$th city, $X_t$ is the GDP per capita ($10^4$ Chinese Yuan) and $\bZ_t$ includes the fiscal expenditure (FE) and fixed assets investment (FAI), which are generally  deemed to be correlated with the industrial structure. To eliminate effect by the difference of economic scales, we divide the FE and FAI by the total GDP for each city, denoted by $Z_{t1}$ and $Z_{t2}$, respectively\footnote[1]{We also separately test the existence of kink effects between $Y_t$ and $Z_{t1}$, $Y_t$ and $Z_{t2}$ at different quantiles. The resulting p-values are  all greater than 0.1  across all quantiles indicating no kink effect on FE and FAI.}. We let $\tau=0.1,0.3,0.5,0.7$ and $0.9$ to study the prefectural-level cities at different development levels.

\begin{table}
\centering
\begin{threeparttable}
\footnotesize
\small
\caption{Parameter estimation and test results of the MKQR model at different quantile levels for secondary industrial structure data of China.}
\label{tab:4}
\begin{tabular}{cccccc}
\hline
\hline
 &  $\tau=0.1$  & $\tau=0.3$ & $\tau=0.5$ & $\tau=0.7$ & $\tau=0.9$ \\
\hline
P-values & 0.000 & 0.000 & 0.000 & 0.000& 0.000\\
$\widehat{K}$ & 1& 1 & 1 & 1 & 1 \\
$\widehat{\alpha}_0$ 	&	$-0.012_{(0.067)}$ 	&	$0.184_{(0.057)}$ 	&	$0.090_{(0.057)}$ 	&	$0.182_{(0.057)}$ 	&	$0.273_{(0.054)}$ 	\\
$\widehat{\alpha}_1$ 	&	$0.114_{(0.022)}$ 	&	$0.059_{(0.014)}$ 	&	$0.102_{(0.019)}$ 	&	$0.084_{(0.017)}$ 	&	$0.074_{(0.015)}$ 	\\
$\widehat{\beta}_1$ 	&	$-0.109_{(0.023)}$ 	&	$-0.054_{(0.014)}$ 	&	$-0.098_{(0.019)}$ 	&	$-0.079_{(	0.017)}$ 	&	$-0.068_{(0.015)}$ 	\\
$\widehat{\delta}_1$	&	$3.457_{(0.307)}$ 	&	$4.490_{(0.355)}$ 	&	$3.468_{(0.227)}$ 	&	$3.571_{(0.191)}$ 	&	$3.777_{(0.301)}$ 	\\
Wald 	&	[2.856,	4.058] 	&	[3.794,	5.184] 	&	[3.023,	3.913] 	&	[3.195,	3.946] 	&	[3.187,	4.367] 	\\
Boot 	&	[2.567,	7.377] 	&	[2.729,	4.831] 	&	[2.599, 4.801] 	&	[3.027, 4.834] 	&	[2.782, 6.552] 	\\
Score &       [2.537, 4.561] &    [2.786, 4.808] &    [3.196, 4.967] &    [2.651, 5.410] &    [3.236, 4.409]\\
$\widehat{\gamma}_1$	&	$-0.985_{(0.405)}$ 	&	$-0.916_{(0.229)}$ 	&	$-0.711_{(0.159)}$ 	&	$-0.828_{(0.199)}$ 	&	$-0.687_{(0.252)}$	\\
$\widehat{\gamma}_2$ 	&	$0.078_{(0.024)}$ 	&	$0.086_{(0.020)}$ 	&	$0.090_{(0.015)}$ 	&	$0.098_{(0.014)}$ 	&	$0.060_{(0.012)}$ 	\\
\hline
\end{tabular}
\begin{tablenotes}
        \footnotesize
       \item The figures in parentheses denote the standard errors of estimators.
      \end{tablenotes}
\end{threeparttable}
\end{table}

 \begin{figure}[!h]
\centering
\includegraphics[scale=0.81]{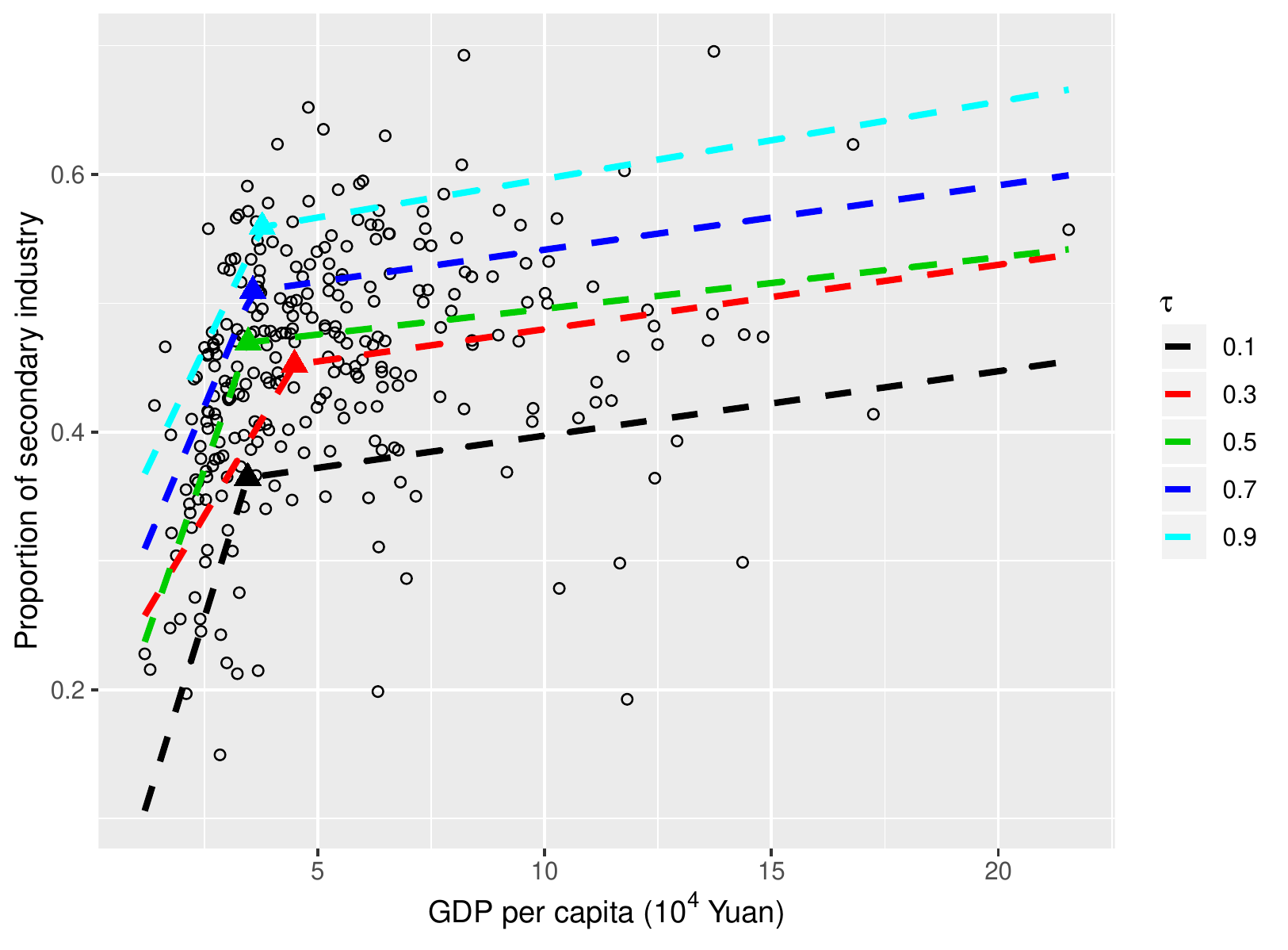}
\caption[]{\label{fig:3} Scatter plot between secondary industrial structure of China and GDP per capita   with the fitted MKQR curves at different quantile levels. $\blacktriangle$ denotes the estimated kink point.}
\end{figure}

Table \ref{tab:4} reports P-values for testing the existence of kink effects based on 1000 bootstrap replicates, the estimated number of kink points, the estimated parameters as well as their the standard errors, and the confidence intervals for kink locations. Figure \ref{fig:3} displays the scatter plot between secondary industrial proportions of 280 cities in China and their GDP per capita with the fitted MKQR curves at different quantile levels. According to Table \ref{tab:4}, P-values are 0 and $\widehat{K}=1$ for all different quantiles, which indicate that there exists a significant kink point.
$\widehat{\alpha}_1>0$ and $\widehat{\beta}_1<0$ for all quantiles are statistically significant
which means that the second industrial proportions $Y_t$ first quickly increase with GDP per capita and then stabilizes with a slow increasing rate of $\widehat{\beta}_1+\widehat{\alpha}_1$ (e.g. $\widehat{\beta}_1+\widehat{\alpha}_1=0.004$ for $\tau=0.5$).
This empirical finding demonstrates the classical economic theory about the process of development.
It is also of interest to observe that the estimated kink points are around 35000 to 45000 Chinese Yuan (roughly 5000-6500 United States Dollar).
Based on the Chenery industrialization stage theory \citep{chenery1986industrialization}\footnote[2]{ Professor Hollis B. Chenery at Harvard University
 believed that modern economic growth can be understood as a comprehensive transformation of the economic structure. He divided the  structural transformation process of GDP per capita into three stages: Initial, Intermediate and Post-industrial stages, corresponding to the GDP per capita less than 1495 dollars, 1495-11214 dollars and greater than 11214 dollars.}, GDP per capita in this interval indicates  that an economic entity is going through an important turning period. During this period, if one economy can skip the threshold value and achieve economic restructuring, it will move into high-income group.  Otherwise, the middle-income trap may loom.
 In addition, both regressors $Z_{t1}$ and $Z_{t2}$ are statistically significant based on the Wald-type test. It is confirmed that the proportions of secondary industry are indeed correlated to the government fiscal expenditure and the fixed assert investment.

\subsection{Triceps Skinfold Thickness of Gambian Females}
Triceps skinfold thickness (TSF) as an important measure for body density experiences the dynamic changes with the increase of age. People whose TSFs are above the 85th percentile are more likely to suffer from obesity, while those whose TSFs are lower the 20th percentile are usually skinny.  Exploring the relationship between TSF and age at different quantiles has been of great interest in biological and human health studies. For instance, \cite{durnin1974body} divided the 481 subjects aged from 16-72 into four subgroups based on the ages and used the linear regression to fit the logarithm of TSF and body densities for each subsample. The results showed that the regression coefficients of each group  exhibited significant differences from the others.  \cite{cole1992smoothing} demonstrated that there existed cubic splines non-linear pattern between the logarithm of TSF and age  by using the smooth fitting curves. \cite{perperoglou2019review} fitted the Gambian females dataset using the spline regression to depict the nonlinearity between TSF and the age. Although the spline regression captures the nonlinear trend, it does not provide any information concerning thresholds and is lack of interpretability in each segment. The nonparametric spline method is either not robust to the outliers and heavy-tailed data.

We consider the dataset collected by \cite{royston2008multivariable} from an anthropometry survey at three Gambian villages in 1989, containing  892 women between the ages of 0 and 55.
To investigate the relationship between their TSFs
and the age and identify the potential kink points at different quantiles, the following MKQR model is considered
\begin{equation}\label{eq:17}
Q_Y(\tau|X_t)=\alpha_0+\alpha_1X_t+\sum_{k=1}^K\beta_k(X_t-\delta_k)I(X_t>\delta_k),\quad t=1,\cdots,892,
\end{equation}
 where $Y_t$ is $\log$(TSF), $X_t$ is the age, $K$ is the unknown number of kink points. We set $\tau =0.1,0.3,0.5,0.7$ and 0.9 to study the different conditional quantiles of $\log$(TSF) on age.

\begin{figure}[!h]
\centering
\includegraphics[scale=0.81]{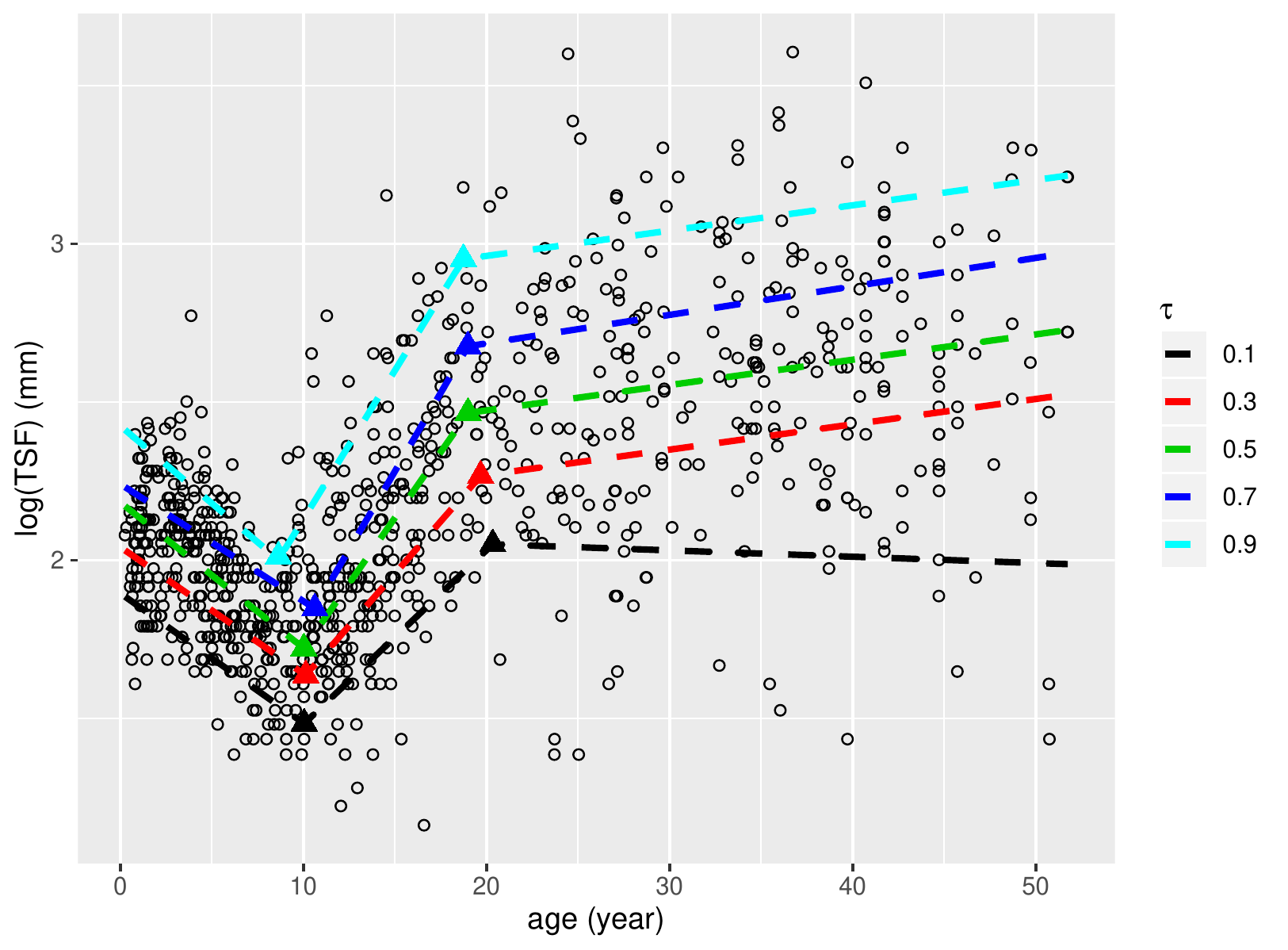}
\caption[]{\label{fig:4}
Scatter plot between the logarithm of TSF and ages for Gambian females  with the fitted MKQR curves at different quantile levels. $\blacktriangle$ denotes the estimated kink point.}
\end{figure}

Table \ref{tab:6} reports P-values for testing the existence of kink effects based on 1000 bootstrap replicates, the estimated number of kink points, the estimated parameters as well as their the standard errors, and the confidence intervals for kink locations.  The resulting P-values are all close to zeros, implying that $\log$(TSF) has significant kink effects on the age for all quantiles. We estimate the MKQR models at different quantiles by setting 10 initial kink points
and identify $\widehat{K}=2$ kink points located round 10 years and 20 years.
This result is in accord with the biological intuition. Two kink points split the domain of the age into the three growth periods of human beings: childhood, adolescence and adults.
Figure \ref{fig:4} also displays the scatter plot between $\log$(TSF) and their GDP per capita with the fitted MKQR curves at different quantile levels. One can observe that the logarithm of TSF decreases quickly with the age in the childhood up to about 8-11 years old, then experiences a growth spurt at adolescence up to about 18-21 years old and finally stays almost stable after then for adults.
The variance of TSF increases with the age, which makes quantile regression necessary
to handle with the heteroscedasticity. It is also interesting to notice that the kink points estimators are heterogeneous across different quantiles. For the higher quantiles such as $\tau=0.9$, $\log$(TSF) tend to experience the smaller kink points locations than other quantiles, for example $\widehat{\delta}_1=8.604$ years for $\tau=0.9$.
It means that Gambian females with obesity reach the biological limits earlier, making their TSFs get changed sooner in the growth process.

 \begin{table}
\centering
\begin{threeparttable}
\footnotesize
\small
\caption{Parameter estimation and test results of the MKQR model at different
quantile levels for triceps skinkfold thickness data for Gambian females.}
\label{tab:6}
\begin{tabular}{cccccc}
\hline
\hline
 &  $\tau=0.1$  & $\tau=0.3$ & $\tau=0.5$ & $\tau=0.7$ & $\tau=0.9$ \\
\hline
P-values & 0.000 & 0.000 & 0.000 & 0.000& 0.007\\
$\widehat{K}$ & 2& 2 & 2 & 2 & 2 \\
$\widehat{\alpha}_0$ &	$1.895_{(0.021)}$ 	&	$2.042_{(0.029)}$ 	& $2.183_{(0.027)}$ & $2.241_{(0.023)}$ & $2.426_{(0.037)}$ \\
$\widehat{\alpha}_1$ &	$-0.041_{(0.002)}$ 	&	$-0.040_{(0.005)}$ 	& $-0.046_{(0.005)}$ &	$-0.037_{(0.003)}$ 	& $-0.049_{(0.008)}$ 	\\
$\widehat{\beta}_1$  &	$0.096_{(0.013)}$ 	&	$0.106_{(0.010)}$ 	& $0.129_{(0.010)}$ 	&$0.136_{(0.012)}$ 	&	$0.144_{(0.015)}$ 	\\
$\widehat{\beta}_2$&	$-0.056_{(0.014)}$ & $-0.058_{(0.010)}$ & $-0.075_{(0.009)}$ 	&	$-0.090_{(0.012)}$ 	&	$-0.086_{(0.013)}$ 	\\
$\widehat{\delta}_1$ &	$10.035_{(0.130)}$ &  $10.117_{(0.379)}$ &  $10.030_{(0.306)}$ &	$10.635_{(0.425)}$ 	&	$8.604_{(0.472)}$ 	\\
Wald&	[9.781, 10.290] &	[9.373, 10.861] &	[9.430, 10.630] &[9.803, 11.467] &[7.679,  9.530] 	\\
Boot &	[8.206, 12.988] &	[9.086, 12.235] &	[9.425, 12.050] &[8.223, 12.202] &[7.737, 10.679]	\\
Score &   [8,217, 13.930] &   [8.979, 12.393] &   [9.418, 12.478] &[7.663, 12.758] &[7.665, 10.967]\\
$\widehat{\delta}_2$ &  $20.414_{(2.927)}$ & $19.689_{(1.525)}$ &	$18.993_{(1.048)}$ 	& $18.964_{(0.845)}$ &	$18.720_{(1.489)}$ 	\\
Wald &	[14.678, 26.150]&	[16.700, 22.678]	&[16.939, 21.047] &[17.307, 20.621]&[15.801, 21.639] 	\\
Boot &	[14.530, 47.470]&	[17.280, 29.735] 	&[17.562, 23.226] &[18.067,	24.724]&[16.588, 24.821] \\
Score  &  [17.487, 49.680] & [16.639, 42.566]  &   [17.945, 25.282] &[18.119, 25.728] & [15.742, 26.166]\\
\hline
\end{tabular}
\end{threeparttable}
\end{table}

 As a comparison, we also analyze the dataset by using SKLS and SKQR methods.  Both methods can only detect with a single kink point at around 6-8 years, which is much lower than our first  threshold estimator $\widehat{\delta}_1$.  However, if the Wald-type test of \cite{li2011bent} and the F-type test of \cite{hansen2017regression} to  check the existence of kink effects using subsmaple in the second segment divided by the threshold estimator,  we find that both tests reject the null hypothesis indicating that some potential kink effect is ignored. In contrast, our MKQR method is flexible and robust in practice to capture multi-kink effects.

\csection{CONCLUSION}
In this article, we studied the flexible multi-kink quantile regression (MKQR) model without knowing the number of kink points. It is robust to outliers and heavy-tailed errors and more
flexible for modelling data with heterogeneous conditional distributions.
We proposed a BRISQ algorithm for estimating parameters. It is much more computationally efficient and not sensitive to the initial values. The selection consistency and the asymptotic normality were
established and the statistical inference for kink effects were also developed.
A R package \emph{MultiKink} has been developed for all the estimation and inference procedures.
Extensions to other regressions such as generalized linear models, Cox proportional hazards models or censored models are relegated to the future research.

\scsection{APPENDIX}
\renewcommand{\theequation}{A.\arabic{equation}}
\renewcommand{\thesection}{A}
\vspace{-0.5cm}
\noindent
\subsection{Proof of Theorem \ref{thm1}:}
 Theorem \ref{thm1} is equvalent to
\begin{equation}\label{the1:eq11}
P\left(\min_{K\neq K_0}sBIC(K)>sBIC(K_0)\right)\rightarrow1.
\end{equation}
To prove \ref{the1:eq11}, we  identify  two different cases i.e Case 1 for $K<K_0$ and Case 2 for $K>K_0$.

\textbf{Case 1:} when $K<K_0$, we need first prove $E\rho_\tau\{Y-Q_Y(\tau;\ttheta_K|\bW)\}>
E\rho_\tau\{Y-Q_Y(\tau;\ttheta_{K_0}|\bW)\}$. From Knight's identity, for any $a,b\in\mathbb{R}$,
$$
\rho_\tau(a-b)-\rho_\tau(a)=b\{I(a\leq0)-\tau\}+\int_0^b\{I(a\leq s)-I(a\leq0)\}ds
$$
We can directly obtain that
\begin{eqnarray*}
&&\rho_\tau\{Y-Q_Y(\tau;\ttheta_K|\bW)\}-\rho_\tau\{Y-Q_Y(\tau;\ttheta_{K_0}|\bW)\}\\
&=& \{Q_Y(\tau;\ttheta_K|\bW)-Q_Y(\tau;\ttheta_{K_0}|\bW)\}\{I(e_{0}\leq0)-\tau\}\\
&&+\int_0^{Q_Y(\tau;\ttheta_K|\bW)-Q_Y(\tau;\ttheta_{K_0}|\bW)}\{I(e_0\leq s)-I(e_0\leq0)\},
\end{eqnarray*}
and therefore
\begin{eqnarray*}
&&E[\rho_\tau\{Y-Q_Y(\tau;\ttheta_K|\bW)\}]-E[\rho_\tau\{Y-Q_Y(\tau;\ttheta_{K_0}|\bW)\}]\\
&=&E\int_0^{Q_Y(\tau;\ttheta_K|\bW)-Q_Y(\tau;\ttheta_{K_0}|\bW)}[F_0(s)-F_0(0)]ds.
\end{eqnarray*}
From Assumption (A1), density value $f_0(\cdot|\bW)$ is always bounded away from zero. We can immediately obtain that $E\int_0^{Q_Y(\tau;\ttheta_K|\bW)-Q_Y(\tau;\ttheta_{K_0}|\bW)}[F_0(s)-F_0(0)]ds>0$ no matter $Q_Y(\tau;\ttheta_K|\bW)-Q_Y(\tau;\ttheta_{K_0}|\bW)$ is positive or negative. Thus, $E\rho_\tau\{Y-Q_Y(\tau;\ttheta_K|\bW)\}>
E\rho_\tau\{Y-Q_Y(\tau;\ttheta_{K_0}|\bW)\}$ holds.

Next, by using similar arguement in Theorem \ref{thm2}, it can be shown that $\|\widehat{\ttheta}-\ttheta\|=O(n^{-1/2})$. Using Knight's identity once again, we have
\begin{eqnarray}\nonumber
&&\sum_{t=1}^n\rho_\tau\{Y_t-Q_Y(\tau;\widehat{\ttheta}_K|\bW_t)\}-\sum_{t=1}^n\rho_\tau\{Y_t-Q_Y(\tau;\ttheta_K|\bW_t)\}\\\nonumber
&=&\sum_{t=1}^n\{Q_Y(\tau;\widehat{\ttheta}_K|\bW_t)-Q_Y(\tau;\ttheta_K|\bW_t)\}\{I(e_t\leq0)-\tau\}\\
&&+\int_0^{Q_Y(\tau;\widehat{\ttheta}_K|\bW_t)-Q_Y(\tau;\ttheta_K|\bW_t)}\{I(e_t\leq s)-I(e_t\leq0)\}ds\label{the1:11}
\end{eqnarray}
By  Assumption (A1)-(A3) along with the fact that $\|\widehat{\ttheta}-\ttheta\|=O(n^{-1/2})$,  taking the expectation for the second term  of (\ref{the1:11}) conditional on $\bW_t$ leads to
\begin{eqnarray}\nonumber
&&\sum_{t=1}^n\int_0^{Q_Y(\tau;\widehat{\ttheta}_K|\bW_t)-Q_Y(\tau;\ttheta_K|\bW_t)}\{F_t(s)-F_t(0)\}ds\\\nonumber
&=&\left(\sum_{t=1}^nf_t(0)\{Q_Y(\tau;\widehat{\ttheta}_K|\bW_t)-Q_Y(\tau;\ttheta_K|\bW_t)\}\right)(1+o(1))=O(n^{1/2}).
\end{eqnarray}
Therefore,
\begin{eqnarray*}
&&\text{sBIC}(K)-\text{sBIC}(K_0)\\
&=&\log\left(1+\frac{\sum_{t=1}^n\rho_\tau\{Y_t-Q_Y(\tau;\widehat{\ttheta}_K|\bW_t)\}/n-\sum_{t=1}^n\rho_\tau\{Y_t-Q_Y(\tau;\ttheta_{K_0}|\bW_t)\}/n}{\sum_{t=1}^n\rho_\tau\{Y_t-Q_Y(\tau;\ttheta_{K_0}|\bW_t)\}/n}\right)\\
&&-O\left(\frac{\log n}{n}C_n\right)\\
&=&\log\left(1+\frac{\sum_{t=1}^n\rho_\tau\{Y_t-Q_Y(\tau;{\ttheta}_K|\bW_t)\}/n-\sum_{t=1}^n\rho_\tau\{Y_t-Q_Y(\tau;\ttheta_{K_0}|\bW_t)\}/n}{\sum_{t=1}^n\rho_\tau\{Y_t-Q_Y(\tau;\ttheta_{K_0}|\bW_t)\}/n}\right.\\
&&+\left.\frac{O(n^{-1/2})}{\sum_{t=1}^n\rho_\tau\{Y_t-Q_Y(\tau;\theta|\bW_t)\}/n}\right)-O\left(\frac{\log n}{n}C_n\right)\\
&\geq&\log\left(1+\frac{O(n^{-1/2})}{\sum_{t=1}^n\rho_\tau\{Y_t-Q_Y(\tau;\theta|\bW_t)\}/n}\right)-O\left(\frac{\log n}{n}C_n\right)\\
&=&O\left(\frac{1}{\sqrt{n}}\right)-O\left(\frac{\log n}{n}C_n\right)
\end{eqnarray*}
where ``$\geq$" is aroused by the fact that $\lim_{n\rightarrow\infty}\sum_{t=1}^n\rho_\tau\{Y_t-Q_Y(\tau;\ttheta_K|\bW_t)\}/n-\sum_t^n\rho_\tau\{Y_t-Q_Y(\tau;\ttheta_{K_0}|\bW_t)\}/n=E\rho_\tau\{Y-Q_Y(\tau;\ttheta_K|\bW)\}-
E\rho_\tau\{Y-Q_Y(\tau;\ttheta_{K_0}|\bW)\}>0$ from the law of large numbers. Therefore $\text{sBIC}(K)-\text{sBIC}(K_0)>0$  for $K<K_0$ when $n$ goes to infinity.

\textbf{Case 2:} i.e. when $K>K_0$, following the similar argument in (\ref{the1:11}), it is easy to show that $|\sum_{t=1}^n\rho_\tau\{Y_t-Q_Y(\tau;\widehat{\ttheta}_K|\bW_t)\}/n-\sum_{t=1}^n\rho_\tau\{Y_t-Q_Y(\tau;\ttheta_{K_0}|\bW_t)\}/n|=O\left(1/n\right)$. We thus have
\begin{eqnarray*}
&&\text{sBIC}(K)-\text{sBIC}(K_0)\\
&=&\log\left(1+\frac{\sum_{t=1}^n\rho_\tau\{Y_t-Q_Y(\tau;\widehat{\ttheta}_K|\bW_t)\}/n-\sum_{t=1}^n\rho_\tau\{Y_t-Q_Y(\tau;\ttheta_{K_0}|\bW_t)\}/n}{\sum_{t=1}^n\rho_\tau\{Y_t-Q_Y(\tau;\ttheta_{K_0}|\bW_t)\}/n}\right)\\
&&+(K-K_0)\frac{\log n}{2n}C_n\\
&=&\log\left(1+\frac{\sum_{t=1}^n\rho_\tau\{Y_t-Q_Y(\tau;{\ttheta}_{K_0}|\bW_t)\}/n-\sum_{t=1}^n\rho_\tau\{Y_t-Q_Y(\tau;\ttheta_{K_0}|\bW_t)\}/n}{\sum_{t=1}^n\rho_\tau\{Y_t-Q_Y(\tau;\ttheta_{K_0}|\bW_t)\}/n}\right.\\
&&\left.+\frac{O(1/n)}{\sum_{t=1}^n\rho_\tau\{Y_t-Q_Y(\tau;\ttheta_{K_0}|\bW_t)\}/n}\right)+(K-K_0)\frac{\log n}{2n}C_n\\
&=&(K-K_0)\frac{\log n}{2n}C_n+O\left(\frac{1}{n}\right).
\end{eqnarray*}
Since $K-K_0>0$, then $\text{sBIC}(K)-\text{sBIC}(K_0)>0$ with probability approaching to one. The proof of Theorem \ref{thm1} is now completed. $\hfill\blacksquare$\\

\subsection{Proof of Theorem \ref{thm2}}
To show the asymptotic normality of $\widehat{\ttheta}$, we need  derive its consistency at first.
\begin{lem}\label{consistency}
Under Assumptions (A1) and (A4)-(A6), $\widehat{\ttheta}$ is  a consistent estimator of $\ttheta_0$.
\end{lem}
\noindent
{\sc Proof of Lemma \ref{consistency}:}
We first need to show that $\sup_{\ttheta\in\Theta}|S_n(\ttheta)-S(\ttheta)|\stackrel{p}{\longrightarrow}0$ as $n\rightarrow\infty$. Notice that $S(\ttheta)$ is continuous and has the following first derivative
$$
\frac{\partial S(\ttheta)}{\partial\ttheta}=\text{E}\left[\psi_\tau\{Y_t-Q_Y(\tau;\ttheta|\bW_t)\}h(\bW_t;\ttheta)\right].
$$
By the Assumptions (A5) and (A6), we can get that $\text{E}\sup_{\ttheta}|h(\bW_t;\ttheta)|<\infty$. Together with $\psi_\tau\{Y_t-Q_Y(\tau;\ttheta|\bW_t)\}\leq\max(\tau,1-\tau)$, we can show that $\text{E}\sup_{\ttheta\in\Theta}\psi_\tau\{Y_t-Q_Y(\tau;\ttheta|\bW_t)\}h(\bW_t;\ttheta)$ is finite. By using the mean-value theorem, for any $\ttheta^1,\ttheta^2\in\Theta$, there exists a $\ttheta^*$ such that
$$
S_n(\ttheta^1)-S_n(\ttheta^2)=\frac{1}{n}\sum_{t=1}^n\left[\psi_\tau\{Y_t-Q_Y(\tau;\ttheta^{*}|\bW_t)\}h(\bW_t;\ttheta^*)\right]\trans(\ttheta^1-\ttheta^2)
$$
By using Assumptions (A5) and (A6) again,
$$
\text{E}\Big|\frac{1}{n}\sum_{t=1}^n\psi_\tau\{Y_t-Q_Y(\tau;\ttheta^*|\bW_t)\}h(\bW_t;\ttheta^*)\Big|\leq B_n<\infty,
$$
where $B_n=\text{E}\sup_{\ttheta\in\Theta}\Big|\max(\tau,1-\tau)h(\bW_t;\ttheta)\Big|$. Hence $B_n=O_p(1)$ and $|S_n(\ttheta^1)-S_n(\ttheta^2)|\leq B_n\|\ttheta^1-\ttheta^2\|$ for every $\mathcal{X}_n$. By applying the Lemma 2.9 of \cite{newey1994large}, we have $\sup_{\ttheta\in\Theta}|S_n(\ttheta)-S(\ttheta)|\stackrel{p}{\longrightarrow}0$ for $\ttheta\in\Theta$.

Since $S_n(\ttheta)$ is continuous w.r.t $\ttheta$, and $S(\ttheta)$ uniquely reaches its global minimum at $\ttheta_0$  (Assumption (A4)), together with $\sup_{\ttheta\in\Theta}|S_n(\ttheta)-S(\ttheta)|\stackrel{p}{\longrightarrow}0$, then we can immediately induce that $\widehat{\ttheta}\stackrel{p}{\longrightarrow}\ttheta_0$ as $n\rightarrow\infty$ by using the Theorem 2.1 of \cite{newey1994large}.$\hfill\blacksquare$ \\

The following lemma is sufficient for deriving the Bahadur representation of $\widehat{\ttheta}$.
\begin{lem}\label{bahadur}
Suppose Assumptions (A1) and (A5)-(A6) hold, for any positive sequence $d_n$ converging to zero, we have
\begin{eqnarray*}
&&\sup_{\parallel\ttheta-\ttheta_0\parallel\leq d_n}\Bigg|\Bigg| n^{-1/2}\sum_{t=1}^n[\psi_\tau\{Y_t-Q_Y(\tau;\ttheta|\bW_t)\}h(\bW_t;\ttheta)-\psi_\tau\{Y_t-Q_Y(\tau;\ttheta_0|\bW_t)\}\\
&&\times h(\bW_t;\ttheta_0)]
-n^{-1/2}E\left[\sum_{t=1}^n\psi_\tau\{Y_t-Q_Y(\tau;\ttheta|\bW_t)\}h(\bW_t;\ttheta)\right]\Bigg|\Bigg|=o_p(1)
\end{eqnarray*}
\end{lem}
\noindent
{\sc Proof of Lemma \ref{bahadur}:}
 Define 
\begin{eqnarray*}
u_t(\ttheta,\ttheta_0)&=&\sum_{k=1}^{K_0+1}\left[\psi_\tau\{Y_t-Q_Y(\tau;\ttheta|\bW_t)\}h(\bW_t;\ttheta)-\psi_\tau\{Y_t-Q_Y(\tau;\ttheta_0|\bW_t)\}\right.\\
&&\left.h(\bW_t;\ttheta_0)\right]\cdot I(\delta_{k-1,0}<X_t<\delta_{k,0})\\
&=&\sum_{k=1}^{K_0+1}u_{t,k}(\ttheta,\ttheta_0)
\end{eqnarray*}
For any $\delta_k\in(\delta_{k-1,0},\delta_{k+1,0})$, $u_{t,k}(\ttheta,\ttheta_0)$ can be partitioned into several parts based on the range of $X_t$,
\begin{eqnarray*}
u_{t,k}(\ttheta,\ttheta_0)&=&u_{t,k}(\ttheta,\ttheta_0)I\{\max(\delta_k,\delta_{k,0})<X_t<\delta_{k+1,0}\}+u_{t,k}(\ttheta,\ttheta_0)\cdot\\
&&I\{\delta_{k-1,0}<X_t\leq\min(\delta_k,\delta_{k,0})\}
+u_{t,k}(\ttheta,\ttheta_0)I(\delta_k\leq X_t<\delta_{k,0})\\
&&+u_{t,k}(\ttheta,\ttheta_0)I(\delta_{k,0}\leq X_t<\delta_k)\\
&=&u_{t,k,1}(\ttheta,\ttheta_0)+u_{t,k,2}(\ttheta,\ttheta_0)+u_{t,k,3}(\ttheta,\ttheta_0)+u_{t,k,4}(\ttheta,\ttheta_0)
\end{eqnarray*}
To prove Lemma \ref{bahadur}, it is sufficient to show $\sup_{\|\ttheta-\ttheta_0\parallel\leq d_n}\|n^{-1/2}\sum_{t=1}^n[u_{t,k,j}-E(u_{t,k,j})]\|=o_p(1)$ for $k=1,\cdots,K_0$ and $j=1,\cdots,4$. The proofs directly follow from the result of Lemma 4.6 in \cite{he1996general}. We only take $u_{t,k,1}(\ttheta,\ttheta_0)$ for illustration and the remaining are the same. For this, we need to check the conditions (B1), (B3) and (B$5^{'}$) in \cite{he1996general}.

The measurability is straightforward for (B1). For (B3), by using mean-value theorem , we have
$$
E_{Y_t}[\|u_{t,k,1}(\ttheta,\ttheta_0)\|^2|\bW_t]\leq Ld_nf^*_t\|\U_t\|^3,
$$
 where $L$ is some positive constant, $\U_t=(1,X_t,\Z_t\trans)\trans$ and $f^*_t$ is some intermediate density satisfying $f^*_t\rightarrow f_t(0)$ almost surely when $n\rightarrow\infty$. It is obvious to obtain (B3). For (B$5^{'}$), let $A_n=L\sum_tf^*_t\|\U_t\|^3$. Under Assumptions (A5) and (A6), we have $A_n=O_p(n)$, and $\max_{1\leq t\leq n}\|u_{t,k,1}(\ttheta,\ttheta_0)\|=O_p(n^{1/2})$. Thus, (B$5^{'}$) is satisfied. By using Lemma 4.6 of \cite{he1996general}, Lemma 2 is therefore established.$\hfill\blacksquare$ \\


\noindent
{\sc Proof of Theorem \ref{thm2}:}
Based on Lemmas \ref{consistency} and \ref{bahadur}, we have
\begin{eqnarray}
&&n^{-1/2}\sum_{t=1}^n\left[\psi_\tau\{Y_t-Q_Y(\tau;\widehat{\ttheta}|\bW_t)\}h(\bW_t;\widehat{\ttheta})-\psi_\tau\{Y_t-Q_Y(\tau;\ttheta_0|\bW_t)\} h(\bW_t;\ttheta_0)\right]\nonumber\\
&&-n^{-1/2}\left[E\sum_{t=1}^n\psi_\tau\{Y_t-Q_Y(\tau;\ttheta|\bW_t)\}h(\bW_t;\ttheta)\right]\Bigg|_{\ttheta=\widehat{\ttheta}}=o_p(1)\label{eq:sup1}
 \end{eqnarray}
Applying the Taylor expansion, we obtain
 \begin{equation}\label{eq:sup2}
\left[E\sum_{t=1}^n\psi_\tau\{Y_t-Q_Y(\tau;\ttheta|\bW_t)\}h(\bW_t;\ttheta)\right]\Bigg|_{\ttheta=\widehat{\ttheta}}=n\D_n(\widehat{\ttheta}-\ttheta_0)+O_p(n(\widehat{\ttheta}-\ttheta_0)^2)
 \end{equation}
where
\begin{eqnarray*}
\D_n&=&n^{-1}\sum_{t=1}^n\frac{\partial E\psi_\tau\{Y_t-Q_Y(\tau;\ttheta|\bW_t)\}h(\bW_t;\ttheta)}{\partial\ttheta}\Big|_{\ttheta=\ttheta_0}\\
&=&n^{-1}\sum_{t=1}^n\frac{\partial([\tau-F_t\{Y_t-Q_Y(\tau;\ttheta|\bW_t)\}]h(\bW_t;\ttheta))}{\partial\ttheta}\Big|_{\ttheta=\ttheta_0}\\
&=&n^{-1}\sum_{t=1}^n\Big([-f_t\{Y_t-Q_Y(\tau;\ttheta_0|\bW_t)\}h(\bW_t;\ttheta_0)h\trans(\bW_t;\ttheta_0)]\\
&&+[\tau-F_t\{Y_t-Q_Y(\tau;\ttheta_0|\bW_t)\}]\frac{\partial h(\bW_t;\ttheta_0)}{\partial\ttheta}\Big)\\
&=&n^{-1}\sum_{t=1}^n\left[-f_t\{Y_t-Q_Y(\tau;\ttheta_0|\bW_t)\}h(\bW_t;\ttheta_0)h\trans(\bW_t;\ttheta_0)\right].
\end{eqnarray*}
Combined with the subgradient condition of quantile regression, we have
\begin{equation}\label{eq:sup3}
n^{-1/2}\sum_{t=1}^n\psi_\tau\{Y_t-Q_Y(\tau;\widehat{\ttheta}|\bW_t)\}h(\bW_t;\widehat{\ttheta})=o_p(1)
\end{equation}
Together with (\ref{eq:sup1}), (\ref{eq:sup2}) and (\ref{eq:sup3}), we have
\begin{eqnarray*}
&&-n^{-1/2}\sum_{t=1}^n[\psi_\tau\{Y_t-Q_Y(\tau;\ttheta_0|\bW_t)\}h(\bW_t;\ttheta_0)]\\
&=&n^{1/2}\D_n(\widehat{\ttheta}-\ttheta_0)+O_p(n^{1/2}(\widehat{\ttheta}-\ttheta_0)^2)+o_p(1).
\end{eqnarray*}
Therefore,
$$
n^{1/2}(\widehat{\ttheta}-\ttheta_0)=-\D_n^{-1}n^{-1/2}\sum_{t=1}^n\psi_\tau\{Y_t-Q_Y(\tau;\ttheta_0|\bW_t)\}h(\bW_t;\ttheta_0)+o_p(1).
$$
By Assumption (A5), it follows that $n^{1/2}(\widehat{\ttheta}-\ttheta_0)$ is asymptotically normal with mean zero and variance matrix $\D^{-1}\C\D^{-1}$, following central limit theorem. This completes the proof of Theorem \ref{thm1}.$\hfill\blacksquare$ \\

\subsection{Proof of Theorem \ref{thm3}}
The following lemma is used for proving the Theorem \ref{thm3}.
\begin{lem}\label{lemma:4}
Under the Assumptions (A1),  (A5)-(A6) and (A8), as  $n\rightarrow\infty$, we have
\begin{itemize}
\item[(\uppercase\expandafter{\romannumeral1})] $\sup_\delta|n^{-1}\sum_{t=1}^n\hat{f}_t(\hat{e}_t)\V_t(X_t-\delta)
I(X_t<\delta)-\bH_1(\delta)|\stackrel{p}{\longrightarrow}0$;
\item[(\uppercase\expandafter{\romannumeral2})] $\sup_\delta|n^{-1}\sum_{t=1}^n\hat{f}_t(\hat{e}_t)\V_t\widehat{\beta}_1(X_t-\delta)
I(X_t>\delta)-\bH_2(\delta,\beta_1)|\stackrel{p}{\longrightarrow}0$;
\item[(\uppercase\expandafter{\romannumeral3})] $\sup_\delta|n^{-1}\sum_{t=1}^n\V_t\V_t\trans\hat{f}_t(\hat{e}_t)-\bH|\stackrel{p}{\longrightarrow}0$.
\end{itemize}
\end{lem}
\noindent
{\sc Proof of Lemma \ref{lemma:4}:}
 We only give the proof for $(\uppercase\expandafter{\romannumeral1})$, since the proof for $(\uppercase\expandafter{\romannumeral2})$ and $(\uppercase\expandafter{\romannumeral3})$ are the same. For $(\uppercase\expandafter{\romannumeral1})$, it is sufficient to show that $\sup_\delta|\widehat{\bH}_1(\delta)-\bH_1(\delta)|=o_p(1)$, where $\widehat{\bH}_1(\delta)=n^{-1}\sum_{t=1}^n\hat{f}_t(\hat{e}_t)\V_t(X_t-\delta)
I(X_t<\delta)$. We have
\begin{eqnarray}\nonumber
&&\widehat{\bH}_1(\delta)-\bH_1(\delta)\\\nonumber
&=&n^{-1}\sum_{t=1}^n\{\hat{f}_t(\hat{e}_t)-f_t(\hat{e}_t)\}\V_t(X_t-\delta)I(X_t<\delta)+\\\nonumber
&&\left\{n^{-1}\sum_{t=1}^nf_t(\hat{e}_t)\V_t(X_t-\delta)I(X_t<\delta)-\bH_{1n}(\delta)\right\}+\left\{{\bH}_{1n}(\delta)-\bH_1(\delta)\right\}\\
&=&(a)+(b)+(c)
\end{eqnarray}

 $\sup_\delta|(a)|=o_p(1)$ holds directly by the uniform convergence property of kernel estimator.  For $(b)$, note that
\begin{equation*}
|(b)|\leq n^{-1}\sum_{t=1}^n\V_t(X_t-\delta)I(X_t<\delta)\max_{1\leq t\leq n}\{f_t(Y_t-\widehat{\aalpha}\trans\V_t)-f_t(Y_t-\aalpha\trans\V_t)\}
\end{equation*}
By our Assumptions (A1), (A5) and (A9), and  $\|\widehat{\aalpha}-\aalpha\|=o_p(n^{-1/2})$ in  Lemma \ref{lemma:5}, together with the mean value theorem, we have
\begin{equation*}
\max_{1\leq t\leq n}\{f_t(Y_t-\widehat{\aalpha}\trans\V_t)-f_t(Y_t-\aalpha\trans\V_t)\}\leq\max_{1\leq t\leq n}\|\V_t\|\cdot|f^{'}(Y_t-\xxi\trans\V_t)|\cdot\|\widehat{\aalpha}-\aalpha\|=o_p(1).
\end{equation*}
where $\xxi$ lies between $\widehat{\aalpha}$ and $\aalpha$. Hence $\sup_\delta|(b)|$ is $o_p(1)$.

Finanlly, for $(c)$, we have $n^{-1}\sum_{t=1}^nf_t(e_t)\V_t(X_t-\delta)I(X_t<\delta)\stackrel{p}{\longrightarrow}Ef_t(e_t)\V_t(X_t-\delta)I(X_t<\delta)=\bH_1(\delta)$ for any given $\delta$ by using  law of large numbers. Then $\sup_\delta|(c)|=o_p(1)$, whose proof  follows the similar line of Lemma 1 in \cite{hansen1996inference} and thus is omitted. Since $(a)$, $(b)$ and $(c)$ are $o_p(1)$ uniformly in $\delta\in\Gamma$, then $\sup_\delta|\widehat{\bH}_1(\delta)-\bH_1(\delta)|=o_p(1)$. The proof is completed.$\hfill\blacksquare$ \\

To assess the power of the proposed kink test, we consider the local alternative model
\begin{equation}\label{eq:12}
Q_Y(\tau|\bW_t)=\alpha_0+\alpha_1X_t+n^{-1/2}\beta_1(X_t-\delta)I(X_t>\delta)+\ggamma\trans\Z_t.
\end{equation}
The following lemma holds.
\begin{lem}\label{lemma:5}
Under Assumptions (A1), (A5)-(A6) and (A8), and the local alternative model (\ref{eq:12}), $\widehat{\aalpha}$ has the following Bahadur representation:
$$
\widehat{\aalpha}-\aalpha_0=\bH^{-1}\left\{n^{-1}\sum_{t=1}^n
\psi_\tau(e_t)\V_t\right\}+n^{-1/2}\bH^{-1}\bH_2(\delta,\beta_1)+o_p(1).
$$
where $e_t=Y_t-\aalpha\trans\V_t-n^{-1/2}\beta_1(X_t-\delta)I(X_t>\delta)$.
\end{lem}
\noindent
{\sc Proof of Lemma \ref{lemma:5}:}
 By using Lemma 4.1 of \cite{he1996general}, we have
\begin{eqnarray*}
&&\sup_{\parallel\aalpha-\aalpha_0\parallel\leq c_n}\Bigg|\Bigg| n^{-1/2}\sum_{t=1}^n\{\psi_\tau(Y_t-\aalpha\trans\V_t)-\psi_\tau(e_t)\}\V_t-\\
&&n^{-1/2}\sum_{t=1}^nE\{\psi_\tau(Y_t-\aalpha\trans\V_t)\V_t|\V_t\}\Bigg|\Bigg|=O_p\left((c_n+n^{-1/2})^{1/2}\log n\right)
\end{eqnarray*}
where $c_n=o(1)$ as $n\rightarrow\infty$. Since $E\{\psi_\tau(Y_t-\aalpha\trans\V_t)|\V_t\}=\tau-F\{(Y_t-\aalpha\trans\V_t)|\V_t\}$, then we can obtain
\begin{eqnarray}\nonumber
&&n^{-1/2}\sum_{t=1}^n[\psi_\tau(Y_t-\aalpha\trans\V_t)-\{\tau-F_t(Y_t-\aalpha\trans\V_t)\}]\V_t\nonumber\\
&=&n^{-1/2}\sum_{t=1}^n\psi_\tau(e_t)\V_t+O_p\left((\parallel\widehat{\aalpha}-\aalpha\parallel+n^{-1/2})^
{1/2}\log n\right).\label{eq:sup4}
\end{eqnarray}
Based on the subgradient condition of quantile regression, we get
$$
n^{-1/2}\sum_{t=1}^n\psi_\tau(Y_t-\aalpha\trans\V_t)\V_t=o_p(1)
$$
Hence,
\begin{eqnarray*}
&&n^{-1/2}\sum_{t=1}^n[\psi_\tau(Y_t-\widehat{\aalpha}\trans\V_t)-\{\tau-F_t(Y_t-\widehat{\alpha}\trans\V_t)\}]\V_t\\
&=&n^{-1/2}\sum_{t=1}^n\{F_t(Y_t-\widehat{\aalpha}\trans\V_t)-\tau\}\V_t+o_p(1)\\
&=&n^{-1/2}\sum_{t=1}^nf_t\{Y_t-{\aalpha_0}\trans\V_t-n^{-1/2}{\beta}_{10}(X_t-\delta_0)I(X_t>\delta_0)\}\V_t\V_t\trans(\widehat{\aalpha}-\aalpha_0)\\
&&-n^{-1}\sum_{t=1}^nf_t\{Y_t-\aalpha_0\trans\V_t-n^{1/2}\beta_{10}(X_t-\delta_0)I(X_t>\delta_0)\}\V_t(X_t-\delta_0)I(X_t>\delta_0)\\
&&+o_p(1)+o_p\left(n^{1/2}(\widehat{\aalpha}-\aalpha_0)\right)\\
&=&n^{1/2}\bH(\widehat{\aalpha}-\aalpha_0)-\bH_2(\delta,\beta_1)+o_p(1)+o_p\left(n^{1/2}(\widehat{\aalpha}-\aalpha_0)\right).
\end{eqnarray*}
Together with (\ref{eq:sup4}), the proof of Lemma \ref{lemma:5} is completed.$\hfill\blacksquare$ \\

\noindent
{\sc Proof of Theorem \ref{thm3}:}
Under the null hypothesis $\beta_1=0$,  $q(\delta,\beta_1)=0$ and thus Theorem \ref{thm3} holds under $H_0$. It remains  to show that Theorem \ref{thm3} holds under $H_1$. By Lemmas \ref{lemma:4} and \ref{lemma:5}, and after some simple algebraic manipulation, it is easy to obtain that
\begin{eqnarray*}
R_n(\delta)&=&n^{-1/2}\sum_{t=1}^n\psi_\tau(Y_t-\widehat{\aalpha}\trans\V_t)(X_t-\delta)I(X_t\leq\delta)\\
&=&n^{-1/2}\sum_{t=1}^n\psi_\tau\{e_t+n^{-1/2}\beta_1(X_t-\delta)I(X_t>\delta)-(\widehat{\aalpha}-\aalpha)\trans\V_t\}(X_t-\delta)I(X_t\leq\delta)\\
&=&n^{-1/2}\sum_{t=1}^n\psi_\tau(e_t)(X_t-\delta)I(X_t\leq\delta)-\bH_1(\delta)\bH^{-1}n^{-1/2}\sum_{t=1}^n\psi_\tau(e_t)\V_t\\
&&-\bH_1(\delta)\bH^{-1}\bH_2(\delta,\beta_1)+o_p(1)\\
&=&n^{-1/2}\sum_{t=1}^n\psi_\tau(e_t)\{(X_t-\delta)I(X_t\leq\delta)-\bH_1(\delta)\bH^{-1}\V_t\}-\bH_1(\delta)\bH^{-1}\bH_2(\delta,\beta_1)+o_p(1)\\
&=&R(\delta)+q(\delta,\beta_1)+o_p(1).
\end{eqnarray*}
The weak convergence of $R(\delta)$ can be obtained directly by  following the proof of \cite{stute1997nonparametric}. This completes the proof of Theorem \ref{thm3}.$\hfill\blacksquare$ \\

\noindent

\subsection{Wild Bootstrap Algorithm for P-Values}
The null asymptotical distribution in Theorem \ref{thm3} can not be directly used for computing the P-values. Instead, we utilize a wild bootstrap procedure to approximate the asymptotically valid P-values. This idea is related to \cite{he2003lack}, \cite{feng2011wild}.

We first introduce the following proposition to give the asymptotic representation for  $R_n(\delta)$, which is easier to compute in practice.
 \begin{prop}\label{proposition:1}
 $R_n(\delta)$ has the asymptotic representation
\begin{equation}
R_n(\delta)=n^{-1/2}\sum_{t=1}^n\omega_t\psi_\tau(v_t)\{(X_t-\delta)I(X_t<\delta)-\bH_{1n}(\delta)\bH_n^{-1}\V_t\}
\end{equation}
where $\{\omega_t;t=1,\cdots,n\}$ is a random sample with zero mean, unit variance, and a finite third moment, and   $\{v_t;t=1,\cdots,n\}$ is independent of $\omega_t$ with $\tau$th quantile zero.
\end{prop}

{\sc Proof of Proposition \ref{proposition:1}:}
 Define
\begin{equation}
R_n^{**}(\delta)=n^{-1/2}\sum_{t=1}^nw_t\psi_\tau(v_t)\{(X_t-\delta)I(X_t\leq\delta)-\bH_{1n}(\delta)\bH_n^{-1}\V_t\}.
\end{equation}
To prove the result, we need to show ($i$) $R_n^*(\delta)$ and $R_n^{**}(\delta)$ are asymptotically equivalent, and ($ii$)  $R_n^{**}(\delta)$ converges to the Gaussian process $R(\delta)$.

For ($i$), it is easy to show that
\begin{eqnarray*}
\sup_{\delta\in\Gamma}\|R_n^*(\delta)-R_n^{**}(\delta)\|&=&\sup_{\delta\in\Gamma}\Big\|n^{-1/2}\sum_{t=1}^nw_t\psi_\tau(v_t)\{\widehat{\bH}_{1n}(\delta)\widehat{\bH}_n^{-1}-\bH_{1n}(\delta)\bH_n^{-1}\}\V_t\Big\|\\
&=&o_p(1),
\end{eqnarray*}
by using Lemma \ref{lemma:4}, along with the consistency of $\widehat{\aalpha}-\aalpha$.

The proof of second part ($ii$) is divided into three steps. Firstly we need to show that the covariance function of $R_n^{**}(\delta)$ converges to that of $R(\delta)$. For any $\delta\in\Gamma$ and $\delta^{'}\in\Gamma$, the covariance function of $R_n^{**}(\cdot)$ is
\begin{eqnarray*}
&& Cov\{R_n^{**}(\delta),R_n^{**}(\delta^{'})\}\\
&=& Cov\Big[n^{-1/2}\sum_{t=1}^nw_t\psi_\tau(v_t)\{(X_t-\delta)I(X_t\leq\delta)-\bH_{1n}(\delta)\bH_n^{-1}\V_t\},\\
&&\quad\quad n^{-1/2}\sum_{t=1}^nw_t\psi_\tau(v_t)\{ (X_t-\delta)I(X_t\leq\delta^{'})-\bH_{1n}(\delta^{'})\bH_n^{-1}\V_t\}\Big]\\
&=& n^{-1}\sum_{t=1}^nCov\Big[w_t\psi_\tau(v_t)\{(X_t-\delta)I(X_t\leq\delta)-\bH_{1n}(\delta)\bH_n^{-1}\V_t \},\\
&&\quad\quad w_t\psi_\tau(v_t)\{(X_t-\delta)I(X_t\leq\delta^{'})-\bH_{1n}(\delta^{'})\bH_n^{-1}\V_t \}\Big]\\
&=&n^{-1}\sum_{t=1}^nE\Big[\{w_t\psi_\tau(v_t)\}^2\{I(X_t\leq\delta)-\bH_{1n}(\delta)\bH_n^{-1}\V_t\}\\
&&\quad\quad\times\{(X_t-\delta)I(X_t\leq\delta^{'})-\bH_{1n}(\delta^{'})\bH_n^{-1}\V_t\}\Big]\\
&\rightarrow&\tau(1-\tau)E\Big[\{(X_t-\delta)I(X_t\leq\delta)-\bH_1\trans(\delta)\bH^{-1}\V_t\}\\
&&\quad\times\{(X_t-\delta^{'})I(X_t\leq\delta^{'})-\bH_1\trans(\delta^{'})\bH^{-1}\V_t\}\Big],\text{almost surely,}
\end{eqnarray*}
by using the fact that $w_t$s are independent of $v_t$s and $E\{w_t\psi_\tau(v_t)\}^2=\tau(1-\tau)$. Obviously,  $Cov\{R_n^{**}(\delta),R_n^{**}(\delta^{'})\}$ is the same as the covariance of $R(\delta)$ in Proposition \ref{proposition:1}.

Next, any finite-dimensional projection of $R_n^{**}(\delta)$ converges to that of the process $R(\delta)$ by the Cramer-Wold device. Finally, note that $\mathcal{F}_n=[\psi_\tau(\cdot)\{(X_t-\delta)I(X_t\leq\delta)-\bH_{1n}(\delta)\bH_n^{-1}\V_t\}:t\in\Gamma]$ is a Vapnik-Chervonenskis (VC) class function of functions. Then we can obtain that $R_n^{**}(\delta)$ is uniformly tight by applying the equicontinuity lemma 15 in \cite{pollard2012convergence}. The proof of Proposition \ref{proposition:1} is now completed.$\hfill\blacksquare$ 

The detailed procedures of wild bootstrap to compute the P-value for testing the existence of kink points  are summarized in Algorithm \ref{alg:3}.
\begin{algorithm}
\caption{~~Wild bootstrap algorithm to compute the P-value. }\label{alg:3}
\begin{algorithmic}
\vspace{0.1cm}
\STATE {\bf Step 1.} Compute $T_n(\tau)$ defined in (\ref{eq:11});
\STATE {\bf Step 2.} \For{b=1:B}{
\STATE {\bf Step 2.1.} Generate an i.i.d sample $\{v_1,\cdots,v_n\}$ from $N(0,1)-\Phi^{-1}(\tau)$;
\STATE {\bf Step 2.2.} Generate an i.i.d sample $\{w_1,\cdots,w_n\}$ that are independent of $v_t$ from two point mass distribution $P(w_t=1)=P(w_t=-1)=0.5$;
\STATE {\bf Step 2.3.} Calculate the quantity $T_{nb}^*(\tau)=\sup_{\delta}|R_n^*(\delta)|$, where
$$
R_n^{*}(\delta)=n^{-1/2}\sum_{t=1}^n\omega_t\psi_\tau(v_t)\{(X_t-\delta)I(X_t<\delta)-\widehat{\bH}_{1n}(\delta)\widehat{\bH}_n^{-1}\V_t\}.
$$
}
\STATE{\bf Step 3.} Calculate the P-value as the proportion of $\{T_{n1}^*(\tau),\cdots,T_{nB}^*(\tau)\}$ exceeding $T_n(\tau)$.
\end{algorithmic}
\end{algorithm}


\subsection{Confidence Intervals for Kink Location Parameters $\ddelta$}
  First, we provide the proof of Proposition \ref{prop1}. Take one kink  model as an illustration.
 To prove Proposition \ref{prop1}, we need to show
$$
\sup_{\ttheta\in\Theta}|\widetilde{{S}}_n(\ttheta)-{S}(\ttheta)|\stackrel{p}{\longrightarrow}0
$$
where $\widetilde{{S}}(\ttheta)=n^{-1}\sum_{t=1}^n\rho_{\tau}\{Y_t-\widetilde{Q}_Y(\tau|\bW_t)\}$ and $S(\ttheta)=E\rho_\tau\{Y_t-Q_Y(\tau|\bW_t)\}$. As the bandwidth $h\rightarrow0$, we have $I(X_t>\delta)=\Phi\left((X_t-\delta)/h\right)+o_p(1)$. Then
\begin{eqnarray*}
&&\sup_{\ttheta\in\Theta}|\widetilde{{S}}_n(\ttheta)-{S}(\ttheta)|\\
&=&\sup_{\ttheta\in\Theta}|\widetilde{{S}}_n(\ttheta)-{S}_n(\ttheta)+{S}_n(\ttheta)-{S}(\ttheta)|\\
&\leq&\sup_{\ttheta\in\Theta}|\widetilde{{S}}_n(\ttheta)-{S}_n(\ttheta)|+\sup_{\ttheta\in\Theta}|{S}_n(\ttheta)-{S}(\ttheta)|\\
&=&(i)+(j).
\end{eqnarray*}
For $(i)$, since $Q_Y(\tau|\bW_t)=\widetilde{Q}_Y(\tau|\bW_t)+\beta_1(X_t-\delta)\left\{I(X_t>\delta)-\Phi\left((X_t-\delta)/h\right)\right\}$, we have
\begin{eqnarray*}
&&\sup_{\ttheta\in\Theta}|\widetilde{{S}}_n(\ttheta)-{S}_n(\ttheta)|\\
&=&\sup_{\ttheta\in\Theta}\Big|n^{-1}\sum_{t=1}^n\left[\rho_\tau\{Y_t-\widetilde{Q}_Y(\tau|\bW_t)\}-\rho_\tau\{Y_t-Q_Y(\tau|\bW_t)\}\right]\Big|\\
&\leq&\sup_{\ttheta\in\Theta}n^{-1}\sum_{t=1}^n\Big|\rho_{\tau}\{Y_t-Q_Y(\tau|\bW_t)\}-\rho_\tau\{\widetilde{Q}_Y(\tau|\bW_t)\}\Big|\\
&=&\sup_{\ttheta\in\Theta}n^{-1}\sum_{t=1}^n\Big|\beta_1(X_t-\delta)\left\{I(X_t>\delta)-\Phi\left((X_t-\delta)/h\right)\right\}\left[I\{Y_t-\widetilde{Q}_Y(\tau|\bW_t)\leq0\}-\tau\right]+\\
&&\int_0^{\beta_1(X_t-\delta)\left\{I(X_t>\delta)-\Phi\left((X_t-\delta)/h\right)\right\}}\left[I\{Y_t-\widetilde{Q}_Y(\tau|\bW_t)\leq s\}-I\{Y_t-\widetilde{Q}_Y(\tau|\bW_t)\leq 0\}\right]ds\Big|\\
&=&o_p(1),
\end{eqnarray*}
where the second equality is by using the Knight's identity. For $(j)$,  by using Uniform Strong Law of Large Numbers (USLLN),  we directly obtain   $\sup_{\ttheta\in\Theta}|{S}_n(\ttheta)-{S}(\ttheta)|=o_p(1)$. Since $(i)$ and $(j)$ are both $o_p(1)$, then $\sup_{\ttheta\in\Theta}|\widetilde{{S}}_n(\ttheta)-{S}(\ttheta)|=o_p(1)$.

Finally, following the similar line of Theorem 2 in \cite{zhang2017composite}, the proof of Proposition \ref{prop1} is completed.
$\hfill\blacksquare$

We detail the procedures to
construct the confidence interval for each kink  parameter by sample splitting in Algorithm \ref{alg:4}.
\begin{algorithm}
\caption{~~Confidence Intervals for Kink Location Parameters $\ddelta$.}\label{alg:4}
\begin{algorithmic}
\STATE {\bf Step 1.} Obtain $\widehat{\ddelta}=(\widehat\delta_1,\cdots,\widehat\delta_{\widehat{K}})\trans$ and $\widehat{K}$ from Algorithm \ref{alg:2} for a given $\tau\in (0,1)$.
\STATE {\bf Step 2.}  \For{$k=1:\widehat{K}$}{
\STATE{\bf Step 2.1} Find the upper bound $\widehat{\delta}_{k,u}$ for $\delta_k$.
\STATE{\bf ~~~Step 2.1.1}
~Test $H_{0k}: \delta_k=\widetilde{\delta}_k^{u}$ for $\widetilde{\delta}_k^{u}=\widehat{\delta}_k+\varrho$, $\varrho$ is a small positive increment.
 \STATE{\bf ~~~Step 2.1.2}
 ~If $H_{0k}$ is not rejected, then let $\widetilde{\delta}^{u}=\widetilde{\delta}^{u}+\varrho$ and  repeat Step 2.2.1.
 ~If $H_{0k}$ is rejected, the upper bound for ${\delta}_k$ is $\widehat{\delta}_{k,u}=\widetilde{\delta}^{u}$.
\STATE{\bf Step 2.2} Find the lower bound $\widehat{\delta}_{k,l}$ for $\delta_k$.
\STATE{\bf ~~~Step 2.2.1}
~Test $H_{0k}: \delta_k=\widetilde{\delta}_k^{l}$ for $\widetilde{\delta}_k^{l}=\widehat{\delta}_k-\varrho$, $\varrho$ is a small positive increment.
 \STATE{\bf ~~~Step 2.2.2}
 ~If $H_{0k}$ is not rejected, then let $\widetilde{\delta}^{l}=\widetilde{\delta}^{l}-\varrho$ and  repeat Step 2.3.1.
 ~~If $H_{0k}$ is  rejected, the lower bound for ${\delta}_k$ is $\widehat{\delta}_{k,l}=\widetilde{\delta}^{l}$.
}
\STATE{\bf Step 3.} The $(1-\alpha)$th confidence interval for $\widehat{\delta}_k$ is $[\widehat{\delta}_{k,l},\widehat{\delta}_{k,u}]$, where
$k=1,2,\cdots,\widehat{K}$.
\end{algorithmic}
\end{algorithm}



\renewcommand{\baselinestretch}{1.00}
\baselineskip=14pt

\bibliographystyle{apalike}
\bibliography{mkqr2}

\end{document}